\documentclass{article}
\usepackage{amssymb}
\usepackage{epsfig}
\usepackage{amsmath}

\input{tcilatex}

\begin{document}

\title{\textbf{Precise relativistic orbits 
in Kerr and Kerr-(anti) de Sitter spacetimes.}}
\author{G. V. Kraniotis \footnote{kranioti@mppmu.mpg.de} \\
Max Planck Institut f$\rm {\ddot u}$r Physik,\\
F$\rm {\ddot o}$hringer Ring 6\\
D-80805 M$\rm {\ddot u}$nchen, \\
Germany \footnote{
MPP-2004-56, May 2004}, \\
}
\maketitle

\begin{abstract}
The timelike geodesic equations resulting from the Kerr gravitational metric 
element are derived and solved exactly including the contribution from 
the cosmological 
constant. The geodesic equations are derived, by solving the Hamilton-Jacobi 
partial differential equation by separation of variables.
The solutions can be applied in the investigation of the motion
of a test particle in the Kerr and Kerr-(anti) de Sitter gravitational 
fields. In particular, we apply the exact solutions of the timelike 
geodesics i) to the precise calculation of dragging (Lense-Thirring effect) 
of a satellite's spherical polar orbit in the gravitational field of Earth 
assuming Kerr geometry,
ii) assuming the galactic centre is a rotating black hole we
calculate the precise dragging of a stellar polar orbit aroung the galactic 
centre for various values of the Kerr parameter including those supported 
by recent observations. The exact solution of non-spherical geodesics in 
Kerr geometry is obtained by using the transformation theory of 
elliptic functions.
The exact solution of spherical polar geodesics 
with a nonzero cosmological constant can be expressed in terms of Abelian 
modular theta functions that solve the corresponding Jacobi's inversion 
problem.

\end{abstract}

\bigskip

\bigskip \newpage

\section{ Introduction}

\subsection{\protect\bigskip Motivation}

Most of the celestial bodies deviate very little from spherical
symmetry, and the Schwarzschild spacetime is an appropriate
approximation for their gravitational field \cite{Karl}. However, 
for some astrophysical bodies the rotation of the mass distribution 
cannot be neglected. A more general spacetime solution of the
gravitational field equations should take this property into account.
In this respect, the Kerr solution \cite{KERR} represents, the curved 
spacetime 
geometry surrounding a rotating mass \cite{OHANIAN}. Moreover, the above 
solution 
is also important  for probing the strong 
field regime of general relativity \cite{REES}.
This is significant, since general relativity has triumphed in large 
scale cosmology \cite{PERLMUTTER, BAHCALL, DEBERNAR,
GVKSBW}, and in predicting solar system effects like the perihelion 
precession of Mercury with a very high precision \cite{Albert, 
Mercury}. 
In a recent paper \cite{Mercury} Kraniotis and Whitehouse, provided 
the exact solution in closed analytic form, of the time-like geodesics 
that describe the motion of a test particle in the  
Schwarzschild gravitational 
field including the contribution from the cosmological constant. The solution for the orbit was expressed in terms of Abelian 
hyperelliptic genus 2 modular functions using the solution  of
Jacobi's inversion problem for hyperelliptic integrals.
For zero cosmological constant the solution was provided by the Weierstra$\ss$ 
Jacobi modular form. Subsequently, the authors applied the solution to the 
compact calculation of the orbit and the perihelion precession of Mercury around the Sun and compared with current experimental data. 
The impressive agreement of the {\em precise theory} developed in 
\cite{Mercury} for the perihelion precession and 
the eccentricity of the orbit, based on the exact solution of the geodesic 
equations, with experiment, the precise determination of the cosmological constant effect in conjuction with the  dedicated efforts of the experimentalists to measure the orbit of Mercury 
with higher accuracy in a series of experiments (BepiColombo ESA
\cite{BepiColombo}, Messenger NASA \cite{Messenger}) motivates the developement of the theory beyond the static nature of the 
 Schwarzschild spacetime. In this way,  one can  determine  all the  
relativistic effects,
and compare the theory  with further measurements of the cosmological and orbital physical parameters.

Furthermore, as was discussed in \cite{Mercury}, the investigation of 
spacetime structures near strong gravitational sources, like neutron stars or 
candidate black hole (BH) systems is of paramount importance for testing the 
predictions of the theory in the strong field regime. The study of geodesics 
are crucial in this respect, in providing information of the 
structure of spacetime in the strong-field limit.  The  generalization of 
the precise theory, to the gravitational field  of a rotating mass, will have important applications also in this domain. 

The study of the geodesics from the Kerr metric are additionally motivated by
recent observational evidence of stellar orbits around the galactic centre, 
which indicates that the spacetime surrounding the $\rm{Sgr\; A^{*}}$ radio source,
which is believed to be a supermassive black hole of 3.6 million solar
masses, is described by the Kerr solution rather than the Schwarzschild 
solution, with the Kerr parameter \cite{GENZEL} 
\begin{equation}
\frac{J}{G M_{BH}/c}=0.52\; (\pm 0.1,\pm 0.08,\pm 0.08)
\label{Galaxy}
\end{equation}
where the reported high-resolution infrared 
observations of $\rm {Sgr\; A^{*}}$
revealed `quiescent' emission and several flares.
This is half the maximum value for a Kerr black hole \cite{SPIN}. 
In the above equation $J$ \footnote{$J=c a$ where $a$ is the Kerr parameter. The interpretation of $ca$ as the angular momentum per unit mass was first given by Boyer and Price \cite{BOPR}. In fact, by comparing with the Lense-Thirring 
calculations \cite{LTPr} they determined the Kerr parameter to be: $a=-
\frac{2 \Omega l^2}{5 c}$, where $\Omega$ and $l$ denote the angular 
velocity and radius of the rotating sphere.}  
denotes the angular momentum of the 
black hole (The error 
estimates here the uncertainties in the period, black hole mass ($M_{BH}$) 
and distance to the galactic centre, respectively; $G$ is the gravitational 
constant and $c$ the velocity of light.)

Taking into account the cosmological constant $\Lambda$ contribution,
the generalization of the Kerr solution is described by the Kerr -de Sitter
metric element which in Boyer-Lindquist (BL) coordinates \footnote{These 
coordinates have the advantage that reduce to the Schwarzschild solution 
with a cosmological constant in the limit $a\rightarrow 0$, see \cite{Boyer}.}
is given by \cite{Zdenek,CARTER}:

\begin{eqnarray}
ds^2&=&\frac{\Delta_r}{\Xi^2 \rho^2}\left(c dt-a\sin^2
\theta d\phi\right)^2
-\frac{\rho^2}{\Delta_r}dr^2-\frac{\rho^2}{\Delta_{\theta}}d\theta^2\nonumber
\\
&-&\frac{\Delta_{\theta}\sin^2
\theta}{\Xi^2 \rho^2}\left(a c dt-(r^2+a^2)d\phi\right)^2
\end{eqnarray}
where 
\begin{eqnarray}
\Delta_r &:=&(1-\frac{\Lambda}{3}r^2)(r^2+a^2)-\frac{2 G M r}{c^2} \nonumber \\
\Delta_{\theta}&:=&1+\frac{a^2 \Lambda}{3}\cos^2 \theta \nonumber \\
\Xi&:=&1+\frac{a^2 \Lambda}{3},\;\;\rho^2:=r^2+a^2  \cos^2 \theta \nonumber \\
 \end{eqnarray}

The above solution is {\em stationary}, but not static.

It is the purpose of this paper, to derive  the
geodesic equations that describe the motion for a test particle in the 
Kerr and Kerr -de Sitter gravitational fields, and obtain 
exact solutions of the corresponding 
equations in an interesting class of possible types of motion, generalizing 
the results of \cite{Mercury}.   

We also apply the exact solutions of the geodesic equations to the following 
situations:

{\em Frame dragging from rotating gravitational mass}: An essential property of the geodesics in Schwarzschild
spacetime is that although the orbit precesses relativistically it remains in the same plane; the Kerr rotation adds longitudinal dragging 
to this precession.  For instance, in the {\em spherical polar 
orbits} we will discuss in 
the main text, (where the particle 
traverses all latitudes, passes through the symmetry axis $z$, infinitely many times) the angle of longitude increases  after a complete oscillation in latitude.  This phenomenon, is in accordance with {\em Mach's principle}.

We shall calculate the dragging of inertial frames in the following situations.
a) Dragging of a satellite's spherical polar orbit in the 
gravitational field of Earth 
assuming Kerr geometry.
b) Dragging of a stellar, spherical polar orbit, in the gravitational field 
of a rotating galactic black hole.

The material of this paper is organized as follows. In section 2, starting 
with the Kerr metric we derive the geodesic equations by integrating 
the Hamilton-Jacobi partial differential equation by separation of variables.
In section 3, we solve exactly the timelike spherical polar geodesics. The 
solution for the orbit is expressed in terms of the Weierstra$\ss$ elliptic 
function. We then apply the exact solution in determing the frame-dragging 
(Lense-Thirring effect) of a satellite's spherical polar orbit in the 
gravitational 
field of Earth as well as the dragging of a stellar spherical polar orbit by a 
galactic black hole. The exact expression of dragging is proportional to the 
real half-period of the Weierstra$\ss$ Jacobi modular form. An alternative 
expression in terms of a hypergeometric function is also provided. In 
section 4,  we solve exactly  spherical non-polar geodesics. In section 5,
we describe the general solution of (non-spherical) timelike geodesics in Kerr metric, using the transformation theory of elliptic functions, an approach 
developed by Abel in \cite{NHA}.
In section 6, we derive the timelike geodesic equations in Kerr metric 
that includes the contribution from the cosmological constant. The derivation 
is produced, by integrating the Hamilton-Jacobi partial differential equation 
again by a separation of variables. We subsequently discuss the  exact 
solution, for spherical 
polar geodesics with a cosmological constant. The integration can be achieved, 
by solving the corresponding
Jacobi's inversion problem for a system of Abelian integrals that we have 
formulated. 
Afterwards, the equatorial geodesis with a cosmological constant 
are presented together with the exact solution of the 
corresponding circular equatorial orbits. 
The application of the 
exact solutions with 
the $\Lambda$ term present will be a subject of a future publication.
Finally, in section 7 we present our conclusions and discuss future research 
directions. In the appendices, we collect some of our formal calculations 
as well as some useful formulas for the solution of cubic and quartic equations and definitions of genus-2 theta functions.

\section{Geodesics and the separability of the Hamilton-Jacobi differential 
equation in Kerr metric}

\subsection{Determination of first geodesic integrals using the Hamilton-Jacobi approach assuming vanishing cosmological constant}

In this section we are going to review the method of  integration of the Hamilton-Jacobi differential equation,  when 
separation of variables is taking place and apply it in deriving the timelike 
geodesic equations that describe the motion of a test particle in a Kerr 
spacetime. The more general case of 
Kerr-de Sitter spacetime is treated in section 6. The method for 
deriving the geodesics, assuming vanishing cosmological constant 
in the Kerr spacetime solution, was first applied  from Carter \cite{carter2}.

The Hamilton-Jacobi differential equation is given by:

\begin{equation}
\frac{\partial{W}}{\partial t}+H\left(t;q_1,q_2,\cdots,q_k;
\frac{\partial W}{\partial q_1},\frac{\partial W}{\partial q_2},\cdots, 
\frac{\partial W}{\partial q_k}\right)=0
\label{RaulJacob}
\end{equation}
where $H$, the characteristic function, depends on the degrees of freedom 
$q_1,q_2,\cdots,q_k,t$ and the momenta $p_i:=\frac{\partial W}{\partial q_i},
i=1,\cdots k$. Also $H$, is a function of second degree in the partial derivatives of $W$. 
Assuming, that  an integral $W(t;q_1,q_2,\cdots,q_k;\alpha_1,\cdots,
\alpha_k)$ has been found,  where $\alpha_1,\cdots,\alpha_k$ are integration constants, 
then the general integral of the differential equations of mechanics 
\begin{eqnarray}
\frac{dq_i}{dt}&=&\frac{\partial H}{\partial p_i}, \nonumber \\
\frac{dp_i}{dt}&=&-\frac{\partial H}{\partial q_i} \nonumber \\
\end{eqnarray}
is provided by

\begin{eqnarray}
\frac{\partial W}{\partial \alpha_1}=\beta_1&,&\;\;\;\;\;p_1=\frac{\partial W}{\partial q_1} \nonumber \\
\cdots &,& \;\;\;\;\;\cdots \nonumber \\
\frac{\partial W}{\partial \alpha_k}=\beta_k &,&\;\;\;\;\;p_k=\frac{\partial W}{\partial q_k} \nonumber \\
\end{eqnarray}
where $\beta_1,\cdots,\beta_k$ denote new arbitrary constants.

An interesting case, which was first studied in great detail by St$\rm {\ddot a}$ckel \cite{Stackel} and Levi-Civita \cite{LEVI}, is when one can integrate the Hamilton-Jacobi differential equation 
through separation of variables. In this case, the characteristic function $W$
takes the form:
\begin{equation}
W=\sum_{i=1}^n W_{i}(q_i)
\label{paul}
\end{equation}
where every term on the right hand side of eq.(\ref{paul}) $W_{i}(q_i)$ depends only on $one$ of the variables $q_i$.

Assuming now, that the characteristic function separates in the Kerr metric 
\begin{equation}
W=W_t(t)+ W_{\phi}(\phi)+W_{\theta}+W_r
\end{equation}
where $W_{\theta},W_r$ are functions of $\theta$ and $r$ coordinates respectively, and then plug this expression in the relativistic Hamiltonian in BL coordinates
\begin{equation}
H(x^{\alpha},p_{\beta})=\frac{1}{2}g^{\mu\nu}p_{\mu}p_{\nu}
\end{equation}
and the relativistic form of eq.(\ref{RaulJacob}), $g^{\mu\nu}\frac{\partial W}{\partial x^{\mu}}\frac{\partial W}{\partial x^{\nu}}+\mu^2=0$,
one gets
\begin{equation}
W=-Ect+\int\frac{\sqrt{R}}{\Delta}dr+\int \sqrt{\Theta}d\theta+L \phi
\label{SepC}
\end{equation}
where
\begin{equation}
\Theta:=Q-\left[a^2(\mu^2-E^2)+\frac{L^2}{\sin^2\theta}\right]\cos^2\theta
\label{poliki}
\end{equation}
and 
\begin{equation}
R:=\left[ (r^2+a^2)E-a L\right]^2-\Delta \left[ \mu^2 r^2+ (L-a E)^2+ Q\right]
\end{equation}
with $\Delta:=r^2+a^2-\frac{2 G M r}{c^2}$. Also $E,L$ are constants of integration associated with the isometries of the Kerr metric. Carter's constant of 
integration is denoted by $Q$.  
Differentiation of (\ref{SepC}), with respect to the integration constants $E,
L,Q,\mu$ leads to the  following set of first-order equations of motion
\cite{carter2}:
\begin{eqnarray}
 \rho^2 \frac{c dt}{d\lambda}&=&\frac{r^2+a^2}{\Delta}P-a\left(a\; E\; \sin^2 \theta-L\right) \nonumber \\
 \rho^2 \frac{dr}{d\lambda}&=&\pm \sqrt{R} \nonumber \\
 \rho^2 \frac{d\theta}{d\lambda}&=&\pm \sqrt{\Theta} \nonumber \\ 
 \rho^2 \frac{d\phi}{d\lambda}&=&\frac{a}{\Delta}P-aE+\frac{L}{\sin^2 \theta} 
\label{kerrgeo}
\end{eqnarray}
where 
\begin{equation}
P:=E (r^2+a^2)-a L
\end{equation}

We proceed now to discuss the exact solution of timelike geodesic 
equations (\ref{kerrgeo}) in closed analytic form, and apply them to the precise determination 
of Lense-Thirring precession of  a satellite's spherical polar orbit in the gravitational 
field of Earth, as well as  in stellar orbits aroung the galactic centre.
We should mention at this point, the extreme black hole solutions $a=1$ 
of spherical non-polar geodesics obtained in \cite{Wilkins} expressed 
in terms of formal integrals. 
The more general case in the presence of the cosmological constant, will be 
discussed in section 6.

\section{ Spherical Polar Geodesics}

Depending on whether or not the coordinate radius $r$ is constant along a 
given timelike geodesic, the corresponding particle orbit is characterized 
as spherical or non-spherical respectively. In this section, we will 
concentrate on spherical polar orbits.

Assuming zero cosmological constant, $r=r_f$, where 
$r_f$ is a constant value and using the last two 
equations of (\ref{kerrgeo}) we  
obtain:
\begin{equation}
\frac{d\phi}{d\theta}=\frac{\frac{a P}{\Delta}-a E+L/\sin^2\theta}{\sqrt{\Theta}}
\label{Polargeod}
\end{equation} 
Equation (\ref{Polargeod}) can be rewritten 
\begin{equation}
d\phi=\frac{d\theta \sin\theta \left\{\frac{a [E (r^2+a^2)-a L]}{r^2+a^2-\frac{2 G M r}{c^2}}-a E+ L/\sin^2 \theta \right\} }{\sqrt{Q(1-\cos^2\theta)-
[a^2 (\mu^2-E^2)(1-
\cos^2\theta)+L^2]\cos^2\theta}}
\end{equation}
Now by defining $z:=\cos^2\theta$, the previous equation can be written as 
follows:
\begin{equation}
d\phi=-\frac{1}{2}\frac{dz}{\sqrt{z^3\alpha-z^2(\alpha+\beta)+Q z}}\times \left\{
\frac{a P}{\Delta}-a E+ \frac{L}{1-z}\right\}
\label{Kugel}
\end{equation}
where 
\begin{equation}
\alpha:=a^2 (\mu^2-E^2),\;\;\beta:=Q+L^2
\end{equation}
It has been shown \cite{StoTsou} that a necessary condition  for an orbit to be 
$polar$ (meaning to intersect the symmetry axis of the 
Kerr gravitational field)  is the vanishing of the parameter $L$, i.e . $L=0$.
Assuming $L=0$, in equation (\ref{Kugel}), we can transform it into the Weierstra$\ss$ form of an elliptic curve by the 
following substitution
\begin{equation}
z:=-\frac{\xi+\frac{\alpha+\beta}{12}}{-\alpha/4}
\label{subwei}
\end{equation}
Thus, we obtain the integral equation
\begin{equation}
\int d\phi=\int -\frac{1}{2}\frac{d\xi}{\sqrt{4 \xi^3-g_2 \xi-g_3}}\times
\left\{
\frac{a P^{\prime}}{\Delta}-a E\right\}
\end{equation}
and this orbit integral can be $inverted$ by the Weierstra$\ss$ modular 
Jacobi form \footnote{For more information on the properties 
of the Weierstra$\ss$ function, the reader is referred to the monographs 
\cite{DON,Silverman}, and the appendix of \cite{GVKSBW}.}
\begin{equation}
\xi=\wp\left(\phi/A\right)
\end{equation} 
where $A:=-\frac{1}{2}\left(\frac{a P^{\prime}}{\Delta}-a E\right), P^{\prime}=E (r^2+ a^2)$.

The Weierstra$\ss$ invariants are given by 
\begin{eqnarray}
g_2 &=&\frac{\alpha^2}{12}+\frac{\beta^2}{12}+\frac{\alpha\beta}{6}-\frac{Q \alpha}{4}=
\frac{1}{12}(\alpha+\beta)^2 -\frac{Q\alpha}{4}\nonumber \\
g_3 &=&\frac{\alpha^3}{216}+\frac{\alpha^2 \beta}{72}+\frac{\alpha \beta^2}{72}
+\frac{\beta^3}{216}-\frac{\alpha^2 Q}{48}-\frac{Q\alpha\beta}{48}=
\frac{1}{216}(\alpha+\beta)^3-
\frac{Q \alpha^2}{48}- \frac{Q\alpha\beta}{48}\nonumber \\
\end{eqnarray}
We note that the discriminant $\Delta^c:=g_2^3-27 g_3^2$ of the 
elliptic curve, vanishes for  
$a=0$, \footnote{in this case $\alpha=0,\beta=Q+L^2 (=Q$ for polar orbits).} 
and this is consistent with the fact that circular orbits in the Schwarzschild 
spacetime do not have modular properties \cite{Mercury}.

Now 
\begin{equation}
\frac{aP^{\prime}}{\Delta}-a E=a \left\{ \frac{E(\frac{r^2}{a^2}+1)}{
\left[ \frac{r^2}{a^2}+1-\frac{2 G M r}{c^2 a^2} \right]}-E \right\}
\end{equation}
Then 
\begin{equation}
d\phi=\frac{dz}{\sqrt{z^3\alpha^{\prime}-z^2 (\alpha^{\prime}+\beta^{\prime})
+Q^{\prime}z}}A^{\prime}
\label{intedif}
\end{equation}
where
\begin{eqnarray}
A^{\prime}&:=&  \left\{ \frac{E(\frac{r^2}{a^2}+1)}{
\left[ \frac{r^2}{a^2}+1-\frac{2 G M r}{c^2 a^2} \right]}-E \right\}
\times\left(-\frac{1}{2}\right) \nonumber \\
&=& \frac{\frac{-E G M r}{c^2 a^2}}{\frac{r^2}{a^2}+1-\frac{2 G M r}{c^2 a^2}}
\end{eqnarray}
where $\alpha^{\prime}:=\frac{\alpha}{a^2},\beta^{\prime}:=Q/a^2=Q^{\prime}$.
Using Eq.(\ref{subwei}) with $\alpha\rightarrow \alpha^{\prime},
\beta \rightarrow \beta^{\prime}$, and integrating Eq.(\ref{intedif})
we get
\begin{equation}
\xi=\wp\left( \phi/A^{\prime}\right)
\label{GVK1}
\end{equation}
with the Weierstra$\ss$ invariants $g_2^{\prime},g_3^{\prime}$ given by 
the expressions
\begin{eqnarray}
g_2^{\prime} &=& \frac{\alpha^{\prime 2}}{12}+\frac{{\beta^{\prime}}^2}{12}+
\frac{\alpha^{\prime}\beta^{\prime}}{6}-\frac{Q^{\prime}\alpha^{\prime}}{4} \nonumber \\
g_3^{\prime} &=& \frac{\alpha^{\prime 3}}{216}+\frac{\alpha^{\prime 2} \beta^{\prime}}{72}+ \frac{\alpha^{\prime} {\beta^{\prime}}^2}{72}+ 
\frac{{\beta^{\prime}}^3}{216}-\frac{{\alpha^{\prime 2}} Q^{\prime}}{48}-\frac{{\alpha^{\prime}\beta^{\prime}} Q^{\prime}}{48}
\end{eqnarray}
Equation (\ref{GVK1}) represents, the 
first exact solution of a spherical polar orbit 
assuming zero cosmological constant,  in closed analytic form, in terms of 
the Weierstra$\ss$ Jacobi modular form.

Equation (\ref{GVK1}) can be rewritten in terms of the original variables 
as 
\begin{equation}
\wp(\phi+\epsilon)=\frac{\alpha^{\prime \prime}}{4}\cos^2 \theta -
\frac{1}{12} \left(\alpha^{\prime \prime}+\beta^{\prime \prime}\right)
\label{satellite}
\end{equation}
where $\alpha^{\prime \prime}:=\alpha^{\prime}/A^{\prime 2},
\beta^{\prime \prime}:=\beta^{\prime}/A^{\prime 2}, Q^{\prime \prime}:=
Q^{\prime}/A^{\prime 2}$ with corresponding Weierstra$\ss$ 
invariants $g_2^{\prime\prime},g_3^{\prime\prime}$. Also 
$\epsilon$ is a constant of integration.

\subsection{Frame dragging in spherical polar geodesics assuming 
vanishing cosmological constant}

We are now going to calculate the Lense-Thirring effect which as we discussed 
in the introduction is the imprint of rotating black holes or rotating mass 
distributions, i.e. the dragging of inertial frames. For this we shall need 
to calculate the periods of the Weierstra$\ss$ modular form.

In order to calculate the periods of the Weierstra$\ss$ function we 
need to determine the roots of the cubic. 
The Weierstra$\ss$ invariants are given in terms of the Kerr parameter $a$ and 
the initial conditions $E,Q$ as follows \footnote{We have set $\mu=1$.}:
\begin{eqnarray}
g_2^{\prime\prime}&=&\frac{1}{12}\frac{\left(a^2(1-E^2)+Q\right)^2}{a^4 A^{\prime 4}}-
\frac{1}{4}\frac{Q(1-E^2)}{a^2 A^{\prime 4}} \nonumber \\
g_3^{\prime\prime}&=&\frac{1}{432a^6 A^{\prime 6}}\left[2 a^6 (1-E^2)^3-3 a^4 (1-E^2)^2 Q-
3 a^2 (1-E^2)Q^2 +2 Q^3\right] \nonumber \\
\end{eqnarray}
while the discriminant $\Delta^c$ of the cubic equation is given by the 
expression
\begin{equation}
\Delta^c=\frac{(1-E^2)^2 Q^2 (a^2(-1+E^2)+Q)^2}{256 a^8 A^{\prime 12}}
\end{equation}
The sign of the discriminant $\Delta^c$ determines the 
roots  of the elliptic curve: $\Delta^c>0$, corresponds
to three real roots  while for $\Delta^c<0$ two roots are complex 
conjugates and the third is real. In the degenerate case $\Delta^c=0$, 
(where at least two roots coincide) the elliptic curve becomes singular and 
the solution is not given by 
modular functions. 
The analytic expressions for  the three roots of the cubic, 
which can be obtained by applying the algorithm of Tartaglia and 
Cardano \cite{TARTACARDA}, are given by

\begin{eqnarray}
e_1 &=& \frac{\left(a^2 (-1+E^2)+2 Q\right)\Delta^2}{12 a^2 E^2 r^2 (GM/c^2)^2} \nonumber \\
e_2 &=& -\frac{\left(2 a^2 (-1+E^2)+Q\right)\Delta^2}{12 a^2 E^2 r^2(GM/c^2)^2} \nonumber \\
e_3 &=& \frac{\left(a^2 (-1+E^2)-Q\right)\Delta^2}{12 a^2 E^2 r^2 (GM/c^2)^2} \nonumber \\
\end{eqnarray}

Since we are assuming spherical orbits, there are two conditions from the 
vanishing of the quartic polynomial $R$ and its first derivative. 
Implementing, these two conditions expressions for the parameter 
$E$ and Carter's constant Q are obtained, 
\begin{eqnarray}
E^2&=&\frac{r \Delta^2}{(r^2+a^2) Z} \nonumber \\
Q&=&\frac{r\left( r^3 G M/c^2+a^2r^2-3 a^2 \frac{GM}{c^2}r+a^4\right)}{Z}-a^2 E^2
\nonumber \\
\label{polarpa}
\end{eqnarray}
where $Z:=r^3+a^2 r-3 \frac{GM}{c^2}r^2+a^2 \frac{GM}{c^2}$ \cite{StoTsou}.

Using the first and fourth line of Eq.(\ref{kerrgeo}) with $L=0$, we 
obtain
\begin{eqnarray}
\frac{c dt}{d\phi}&=&\frac{\frac{r^2+a^2}{\Delta}P^{\prime}-a^2 E \sin^2 \theta}{\frac{a}{\Delta}P^{\prime}-aE} \nonumber \\
&=&a+\frac{\frac{r^2 P^{\prime}}{\Delta}}{\frac{aP^{\prime}}{\Delta}-aE}+
\frac{a^2 E \cos^2\theta}{\frac{aP^{\prime}}{\Delta}-aE} 
\end{eqnarray}
or 
\begin{equation}
ct+{\cal E}=a\phi+\frac{r^2 P^{\prime}/\Delta}{(-2 a A^{\prime})}\phi-
\frac{4a^2 E/ \alpha^{\prime\prime}}{(-2 a A^{\prime})}
\left( \zeta(\phi)-\frac{1}{12}(\alpha^{\prime\prime}+\beta^{\prime \prime})\phi \right)
 \end{equation}
where we used Eq.(\ref{satellite}) and the fact that, 
$\int\wp(\phi)d\phi=-\zeta(\phi)$, where $\zeta(z)$ denotes the 
Weierstra$\ss$ zeta function. Also ${\cal E}$ denotes a constant of integration.

Our free parameters are $a$ and $r$. For a given choice of our free parameters,
$E$ and Carter's constant $Q$ are fixed with the aid of (\ref{polarpa}).

The two half-periods $\omega$ and $\omega^{\prime}$ are given by the 
following Abelian integrals (for $\Delta^c>0$) \cite{WHITAKKER}:
\begin{equation}
\omega=\int_{e_1}^{\infty}\frac{dt}{\sqrt{4t^3-g_2t-g_3}}, 
\;\;\;\omega^{\prime}=i\int_{-\infty}^{e_3}\frac{dt}{\sqrt{-4t^3+g_2t+g_3}}
\end{equation}
The values of the Weierstra$\ss$ function at the half-periods \footnote{
An alternative expression for the real half-period $\omega$ of 
the Weierstra$\ss$ function is:
$\omega=\frac{1}{\sqrt{e_1-e_3}}\frac{\pi}{2}F(\frac{1}{2},\frac{1}{2},1,
\frac{e_2-e_3}{e_1-e_3})$, where $F(\alpha,\beta,\gamma,x)$ is the hypergeometric function $1+\frac{\alpha.\beta}{1.\gamma}x+\frac{\alpha(\alpha+1)\beta(\beta+1)}{1.2.\gamma(\gamma+1)}x^2+\cdots$.}
are 
the three roots of the cubic.
For positive discriminant $\Delta^c$ one half-period is real while the 
second is imaginary \footnote{We organize the roots as:$\;\;e_1>e_2>e_3.$}.
The period ratio is defined as $\tau=\frac{\omega^{\prime}}{\omega}$.

After a complete oscillation in latitude, the angle of longitude, 
which determines the amount of dragging for the spherical 
polar orbit in the General Theory of 
Relativity (GTR), increases by 
\begin{equation}
\Delta\phi^{{\rm GTR}}=4 \omega
\end{equation}

\subsubsection{Dragging of  satellite polar orbit in the gravitational field of rotating Earth} 

In this subsection, assuming the Earth's gravitational field is described by a Kerr space-time, we apply the exact solution of the Kerr spherical polar 
geodesics obtained in the previous section, in determining the effect of Earth's rotation on the 
motion of its orbiting satellites. 
Thus we perform  a {\em precise calculation} of the corresponding 
Lense-Thirring effect.

Experimentally, a 
polar orbital geometry has the advantage that classical nodal precessions 
due to the even zonal harmonics of the geopotential, are vanishing. 
We note, that the Gravity Probe B (GP-B) mission, launched in April 2004, retains a stricly polar orbital configuration during the 
science phase \cite{Everitt}, \footnote{http://einstein.stanford.edu/}
 \footnote{Indeed if the orbit is exactly polar the precession due to Earth's quadrupole moment $J_2$, vanishes. Practically, it had originally been suggested by Van Patten and Everitt a two satellite experiment, with the two satellites in similar nearly polar orbits. The quadrupole moment of the Earth, then produces precessions of opposite directions for the two satellites while the Lense-Thirring precession produces the same effect for both.  
To separate these two effects, the small angle between their orbits needs to be measured in addition to the precession rates of the two satellites.
The distance of the two satellites 
at each polar crossing is measured by Doppler ranging. 
Substracting, the contribution of the quadrupole moment from the actual 
measurement determines the Lense-Thirring effect.
Ciufolini \cite{CIU} 
alternatively suggested,  to make use of a satellite already in orbit LAGEOS and require only one new satellite 
LAGEOS II. Although the orbirt of LAGEOS, is far from polar, if the orbit of 
LAGEOS II is inclined at an exactly opposite angle with respect to the axis 
of the Earth, the Lense-Thirring effect can directly be measured, since the contributions from the quadrupole moment cancel exactly. See also \cite{Iorio} 
for a different approach.}.

The Kerr parameter that corresponds to the  angular momentum of the Earth is 
equal to $a_{\bigoplus}=329.432\;{\rm cm}=371.398(2GM_{\bigoplus}/c^2)$.
For the radius of our spherical orbit we use the semi-major axis of the 
GP-B mission $r=r_{{\rm GP-B}}=7027.4\;{\rm km}$ \footnote{We note the 
small eccentricity (0.014) of the GP-B satellite.} . Then Carter's constant $Q$ and the parameter $E$ are determined by eq.(\ref{polarpa}). Then, the half-period $\omega$ 
has the value $\omega=3.699746\times 10^{-11}$, which leads to 
the {\em precise} frame-dragging effect of 
$\Delta_{\phi}^{{\rm GTR}}=0.164\;{\rm \;arcs\;yr^{-1}}$.
We repeated the analysis, for slight different values of radius $r$, and for 
fixed Kerr parameter. 
In particular we used as a value for $r$ the semi-major axis of the 
Polares satellite, 
i.e. $r=r_{{\rm Polares}}=8378\;$km \cite{IORIO2}.
We list our results in table {\ref{EINSTEIN}. 
We also compare our precise results with the Lense-Thirring (LT) 
formula \cite{LTPr}:
\begin{equation}
\Delta\phi^{{\rm LT}}=\frac{2 G J}{c^2r_{\bf e}^3  (1-{\bf e}^2)^{3/2}}
\label{PHATHI}
\end{equation}
in which $J$ is the magnitude of the central body's angular momentum, 
$r_{\bf{e}}$ the semi-major axis of the orbit and ${\bf e}$ the eccentricity 
of the orbit.
For zero eccentricity, formula (\ref{PHATHI}) predicts 
$0.164\;\rm{arcs\;yr^{-1}}$ and $0.096\;\rm{arcs\;yr^{-1}}$  
for the radii of GP-B mission and Polares satellites, respectively.

\begin{table}
\begin{center}
\begin{tabular}{|c|c|c|}\hline\hline
{\bf parameters} & {\bf half-period} & {\bf predicted dragging} \\

$a_{\bigoplus}=329.432$cm, $r=r_{{\rm GP-B}}=7027.4\;$km & $\omega=3.699746\times 
10^{-11}$ &  $\Delta \phi^{{\rm GTR}}=0.164{\rm
\frac{arcs}{yr}}$ \\
$a_{\bigoplus}=329.432$cm, $r=r_{{\rm POLARES}}=8378\;$km 
& $\omega=2.8421923\times 10^{-11}$ & 
$ \Delta \phi ^{{\rm GTR}}=0.0969{\rm
\frac{arcs}{yr}}$ \\
\hline \hline
\end{tabular}
\end{center}
\caption{Predictions for frame dragging of a satellite's polar spherical 
orbit in the gravitational field of Earth. The period ratio 
is $\tau=6.18642 i$.}
\label{EINSTEIN}
\end{table}


\subsubsection{Precise calculation of dragging of stellar 
polar orbits from a galactic black hole}

Assuming, that the centre of the Milky Way is a black hole and that the 
 structure of spacetime  near the region $\rm {Sgr\;A^{*}}$, is described by 
the Kerr geometry as is indicated by observations Eq.(\ref{Galaxy}), 
we determined the precise frame dragging (Lense-Thirring effect) of a stellar orbit with a  spherical polar
geometry. The results are displayed in table {\ref{EINSTEIN1}}.


\begin{table}
\begin{center}
\begin{tabular}{|c|c|c|}\hline\hline
{\bf parameters} & {\bf half-period} & {\bf predicted dragging} \\

$a_{{\rm Galactic}}=0.52, r=10$ & $\omega=0.0515693$ &  
$\Delta \phi^{{\rm GTR}}=11.8188^{\circ} {\rm \;per\; revolution}=42531.4\frac{
{\rm arcs}}{{\rm revolution}}$ \\
$a_{{\rm Galactic}}=0.52, r=50$ 
& $\omega=0.00462023$ & 
$ \Delta \phi ^{{\rm GTR}}=1.0589^{\circ}{\rm
\;per\;revolution}=3810\frac{{\rm
arcs}}{{\rm revolution}}$ \\
\hline \hline
\end{tabular}
\end{center}
\caption{Predictions for frame dragging from galactic black hole, with 
Kerr parameter $a_{{\rm Galactic}}=0.52\frac{GM_{BH}}{c^2}$, for two different 
values of orbital radius. The values of the radii are in units of $GM_{BH}/c^2$ The period ratios, $\tau$, are $2.92558$i and $3.81583$i respectively.}
\label{EINSTEIN1}
\end{table}

We repeated the analysis for a value of the Kerr parameter as high as 
$a_{{\rm Galactic}}=0.9939$. 
Such high values for the angular momentum of the 
black hole, have been recently reported  from X-ray flare analysis of the 
galactic centre \cite{Porquet}.
The results are given in table \ref{EINSTEIN2}.

\begin{table}
\begin{center}
\begin{tabular}{|c|c|c|}\hline\hline
{\bf parameters} & {\bf half-period} & {\bf predicted dragging} \\

$a_{{\rm Galactic}}=0.9939, r=10$ & $\omega=0.0981121$ &  $\Delta \phi^{{\rm GTR}}=22.485
^{\circ} {\rm\;per \;revolution}=80917.2\frac{{\rm
arcs}}{{\rm revolution}}$ \\
$a_{{\rm Galactic}}=0.9939, r=50$ 
& $\omega=0.00882902$ & 
$ \Delta \phi ^{{\rm GTR}}=2.023^{\circ}{\rm
per\;revolution}=7281.66\frac{{\rm
arcs}}{{\rm revolution}}$ \\
\hline \hline
\end{tabular}
\end{center}
\caption{Predictions for frame dragging from galactic black hole, with 
Kerr parameter $a_{{\rm Galactic}}=0.9939\frac{GM_{BH}}{c^2}$, for two different 
values of orbital radius. The values of the radii are in units of $GM_{BH}/c^2$.The period ratios, $\tau$, are $2.5086$ i, $3.40336$ i respectively.}
\label{EINSTEIN2}
\end{table}

\section{Spherical geodesics with $L \not = 0 $}

We now integrate Eq.(\ref{Kugel}) including the contribution 
from the parameter $L$.
Let us define:
\begin{equation}
\Pi:=\int\frac{d\xi}{\sqrt{4 \xi^3- g_2 \xi-g_3}}
\end{equation}
thus $\xi=\wp(\Pi+\epsilon)$, and $A^{\prime \prime} \int\frac{dz}{\sqrt{z^3 \alpha-
z^2 (\alpha+\beta)+Qz}}= A^{\prime \prime}\int\frac{d\xi}{\sqrt{4 \xi^3- g_2 \xi-g_3}}=
A^{\prime \prime}\Pi,\; A^{\prime \prime}:= -\frac{1}{2}\left(\frac{a P}{\Delta}-a E\right)$ and $\epsilon$ is a constant of integration.
Now
\begin{equation}
-\frac{1}{2} L \int \frac{dz}{(1-z)\sqrt{z^3 \alpha-
z^2 (\alpha+\beta)+Q z}}
\end{equation}
under the substitution (\ref{subwei}) becomes 
\begin{eqnarray}
&-&\frac{L \alpha}{8}\int \frac{d\xi}{\left(\frac{\alpha}{4}(1-
\frac{\alpha+\beta}{3\alpha})-\xi\right)\sqrt{4 \xi^3-g_2 \xi-g_3}} \nonumber \\
&=&-\frac{L \alpha}{8}\int \frac{d\xi}{\left(w-\xi\right)\sqrt{4 \xi^3-g_2 \xi-g_3}} \nonumber \\
&=&- \frac{L \alpha}{8}\int \frac{\wp^{\prime}(\Pi)d \Pi}{\left(w-
\wp(\Pi+\epsilon)\right)\sqrt{4 \wp^3(\Pi)-g_2 \;\wp(\Pi)-g_3}} \nonumber \\
&=&- \frac{L \alpha}{8}\int \frac{d\Pi}{\left(w-\wp(\Pi+\epsilon)
\right)} \nonumber \\
&=&- \frac{L \alpha}{8}\left[{\rm Log} \frac{\sigma(\Pi+\epsilon-v_0)}{
\sigma(\Pi+\epsilon+v_0)}+2 \Pi\; \zeta(v_0)\right]\times \frac{1}{\wp^{\prime}
(v_0)} \nonumber \\
\end{eqnarray}
where $w:=\frac{\alpha}{4}\left(1-\frac{\alpha+\beta}{3\alpha}\right) = \wp(v_0)$.
Also $\sigma(z)$ denotes the Weierstra$\ss$ sigma function.
Thus the equation for the orbit is given by
\begin{equation}
\phi=\int d\phi=A^{\prime \prime}\Pi- \frac{L \alpha}{8}\left[{\rm Log} \frac{\sigma(\Pi+\epsilon-v_0)}{
\sigma(\Pi+\epsilon+v_0)}+2 \Pi \zeta(v_0)\right]\times \frac{1}{\wp^{\prime}
(v_0)}
\end{equation}
and $\wp^{2 \prime}(v_0)=4 \wp^3(v_0)-g_2 \wp(v_0)-g_3=4 w^3-g_2 w-g_3$.
In terms of the integration constants, $w$ is given by the expression:
\begin{equation}
w=\frac{a^2(1-E^2)}{4}-\frac{a^2(1-E^2)+Q+L^2}{12}
\end{equation}

Similarly using the first and third line of Eq.(\ref{kerrgeo}), we obtain for $t$ the expression
\begin{equation}
c\; t=\frac{r^2+a^2}{\Delta}P\frac{\Pi}{-2}+\frac{a\Pi}{2}(aE-L)+
\frac{a^2E}{2}\Pi\left(-\frac{1}{3}\frac{\alpha+\beta}{\alpha}\right)+
\frac{a^2 E}{2}\frac{4}{\alpha}\zeta(\Pi)
\end{equation}

\section{General solution for the time-like geodesics in Kerr metric}

In the general case, of non-spherical orbits one has to solve the differential equations
\begin{equation}
\int^\theta \frac{d\theta}{\sqrt{\Theta}}=\int^r \frac{dr}{\sqrt{R}}
\label{getimelike}
\end{equation}
where $R(r)$ is a quartic polynomial 
given by
\begin{equation}
R=\left[ (r^2+a^2)E-a L\right]^2-\Delta \left[ r^2+ (L-a E)^2+ Q\right]
\end{equation}
and $\Theta(\theta)$ is given 
by equation (\ref{poliki}). Note that this equation is an equation 
between two elliptic integrals, in the general case when 
all roots are distinct. The left hand side can be transformed into 
an elliptic integral with variable $z$ or $\xi$ in Weierstra$\ss$ normal form, 
see Eq.(\ref{subwei}).
In order to solve this differential equation and determine $r$ as a function 
of $\theta$, we will employ the theory 
of Abel for the transformation of elliptic functions \cite{NHA}.

The general case of non-spherical orbits is of importance for determining 
the precession of $perihelion,periapsis$, and $perinigricon$ (point of 
closest approach to the black hole).

\subsection{Exact solution of the general Kerr geodesic using the transformation 
theory of elliptic functions}

Abel in \cite{NHA} deals with the following question:

`` Find all possible cases in which the differential equation

\begin{equation}
\frac{dy}{\sqrt{(1-c_1^2 y^2)(1-e_1^2 y^2)}}=\pm C \frac{dx}{\sqrt{(1-c^2 x^2)(1-e^2 x^2)}},
\label{TRANSF}
\end{equation}

is satisfied by putting $y$ an algebraic function of $x$, rational or irrational. ''

He explains that the problem may be reduced to the case in which $y$ is a rational function of $x$. Below we shall describe his approach, which uses the idea 
of $inversion$ of elliptic integrals and then we shall apply it to the solution of eq.(\ref{getimelike}).
Abel's notation is as follows:
From the elliptic integral 
\begin{equation}
{\vartheta}=\int_0 \frac{dx}{\sqrt{(1-c^2 x^2)(1-e^2 x^2)}},
\label{funda}
\end{equation}
he defines the inverse function
\begin{equation}
x:=\lambda \vartheta
\end{equation}
while 
\begin{equation}
\frac{\omega}{2}=\int_0^{1/c}\frac{dx}{\sqrt{(1-c^2 x^2)(1-e^2 x^2)}},\;
\frac{\omega^{\prime}}{2}=\int_0^{1/e}\frac{dx}{\sqrt{(1-c^2 x^2)(1-e^2 x^2)}}
\end{equation}
and $\Delta\vartheta=\sqrt{(1-c^2 x^2)(1-e^2 x^2)}$.
Also the solution of the equation $\lambda\vartheta^{\prime}=\lambda\vartheta$ 
is $\vartheta^{\prime}=(-1)^{m+m^{\prime}}\vartheta+m\omega+m^{\prime}\omega^{\prime}$
where $m,m^{\prime}$ are arbitrary integral numbers.

Let $y=\psi(x)$ be the rational functions we are looking for and $x=
\lambda{\vartheta},x_1=\lambda \vartheta_1$ two roots of the equation 
$y=\psi(x)$, ($y=\psi(x)=\psi(x_1)$). 
Let the radical on the left hand side of Eq.(\ref{TRANSF}), be denoted by $\sqrt{R}$, then
\begin{equation}
\frac{dy}{\sqrt{R}}=\pm C d\vartheta
\end{equation}
Then by changing $x$ with $x_1$, we have $\pm\frac{dy}{\sqrt{R}}=
\pm C d\vartheta_1$, which gives $d\vartheta_1=\pm d\vartheta$ or 
after integration $\vartheta_1=\epsilon \pm \vartheta$ where $\epsilon$ is 
a constant.
The quantities $\lambda \vartheta$ and $\lambda (\vartheta+\epsilon)$ are 
roots, thus we have
\begin{equation}
y=\psi\left(\lambda \vartheta\right)=\psi\left(\lambda (\vartheta+\epsilon)\right)
\end{equation}
Now
\begin{equation}
y=\psi\left(\lambda \vartheta\right)=\psi\left(\lambda (\vartheta+\epsilon)\right)=\psi\left(\lambda(\vartheta+2\epsilon)\right)=\cdots= \psi\left(\lambda(\vartheta+k\epsilon)\right)
\end{equation}
where $k$ denotes an arbitrary integral number.
As the equation $y=\psi(x)$ has only a finite number of roots, there exist 
$k$ and $k^{\prime}$ distinct such that $\lambda(\vartheta+k\epsilon)=\lambda
(\vartheta+k^{\prime}\epsilon)$. Assuming $n:=k-k^{\prime}>0$, and 
replacing $\vartheta$ by $\vartheta-k^{\prime}\epsilon$, we get 
\begin{equation}
\lambda(\vartheta+n\epsilon)=\lambda\vartheta
\end{equation}
Thus we get $\vartheta+n\epsilon=(-1)^{m+m^{\prime}}\vartheta+m\omega+m^{\prime}\omega^{\prime}$, and since $\vartheta$ is a variable we get $(-1)^{m+m^{\prime}}=1$ and
\begin{equation}
\epsilon=\frac{m}{n}\omega+\frac{m^{\prime}}{n}\omega^{\prime}
\label{SYNARTISI}
\end{equation}
where $\mu:=m/n,\mu^{\prime}:=m^{\prime}/n \in Q$, $m+m^{\prime}$ is even number.
Assuming the degree of the equation $y=\psi(x)$ is such that it has roots 
other then $\lambda(\vartheta+k\epsilon)$, anyone of them has the form 
$\lambda(\vartheta+\epsilon_1)$ where $\epsilon_1=\mu_1\omega+
\mu_1^{\prime} \omega^{\prime}, (\mu_1,\mu_1^{\prime}\in Q)$ and all the 
$\lambda(\vartheta+k\epsilon+k_1\epsilon_1)$ are roots of the equation.
Continuing in this fashion the 
roots of $y=\psi(x)$ are of the form
\begin{equation}
x=\lambda(\vartheta+k_1\epsilon_1+k_2\epsilon_2+k_3\epsilon_3+\cdots+
k_{\nu}\epsilon_{\nu})
\end{equation}
where $k_1,k_2,\cdots k_{\nu}$ are integers and the quantities 
$\epsilon_1,\epsilon_2,\cdots \epsilon_{\nu}$ are of the form 
$\mu \omega+\mu^{\prime} \omega^{\prime}$.
By denoting the values of the expression $\lambda(\vartheta+
k_1\epsilon_1+k_2\epsilon_2+\cdots k_{\nu} \epsilon_{\nu})$ by 
$\lambda\vartheta,\lambda(\vartheta+\epsilon_1),\lambda(\vartheta+\epsilon_2),
\cdots \lambda(\vartheta+\epsilon_{ m -1})$, and writing 
$\psi(x)=\frac{p}{q}$, where $p,q$ are polynomials of degree $m$ in $x$, with no common divisor, the equation $y=\psi(x)$ is written
\begin{equation}
p-qy=(f-gy)(x-\lambda\vartheta)(x-\lambda(\vartheta+\epsilon_1))\cdots(
x-\lambda(\vartheta+\epsilon_{m-1}))
\end{equation}
where the constants $f,g$ are the dominant coefficients of $p,q$ respectively.
If $f^{\prime}$ and $g^{\prime}$ denote the coefficients of $x^{m-1}$ we have that
\begin{equation}
f^{\prime}-g^{\prime}y=-(f-gy)\left[\lambda\vartheta+
\lambda(\vartheta+\epsilon_1)+\lambda(\vartheta+\epsilon_2)+\cdots 
\lambda(\vartheta+\epsilon_{m-1})\right]
\end{equation}
and if we define
\begin{equation}
\varphi\vartheta:=\lambda\vartheta+\lambda(\vartheta+\epsilon_1)+
\lambda(\vartheta+\epsilon_2)+\cdots \lambda(\vartheta+\epsilon_{m-1})
\end{equation}
we get for $y$ the expression
\begin{equation}
y=\frac{f^{\prime}+f.\varphi\vartheta}{g^{\prime}+g.\varphi\vartheta}
\end{equation}
By hypothesis, $y$ is a rational function in $x$ and so $\varphi$ must be 
the same. The goal is to express $\varphi$ rationally in $x$ and to determine
$f,f^{\prime},g,g^{\prime},e_1,c_1$ and $C$ in order that (\ref{TRANSF}) is 
satisfied.

First, we search for values of $\epsilon$ which satisfy $\lambda(\vartheta-
\epsilon)=\lambda(\vartheta+\epsilon)$, that is to say $\lambda(\vartheta+2\epsilon)=\lambda\vartheta$. According to (\ref{SYNARTISI}), 
\begin{equation}
\epsilon=\frac{m}{2}\omega+\frac{m^{\prime}}{2}\omega^{\prime}
\end{equation}
where $m+m^{\prime}\in 2Z$. In this case $\lambda(\vartheta+\epsilon)$ takes 
the following distinct values \cite{NHA}
\begin{eqnarray}
\lambda\vartheta&=&x, \nonumber \\
\lambda(\vartheta+\omega)&=&-\lambda\vartheta=-x, \nonumber \\
\lambda(\vartheta+\frac{\omega}{2}+\frac{\omega^{\prime}}{2})&=&
-\frac{1}{ec}\frac{1}{x}, \nonumber \\
\lambda(\vartheta+\frac{3\omega}{2}+\frac{\omega^{\prime}}{2})&=&-\frac{1}{ec}\frac{1}{\lambda(\vartheta+\omega)}=\frac{1}{ec}\frac{1}{x} \nonumber \\
\end{eqnarray}
All the other roots $\lambda(\vartheta+\epsilon)$ come with a term 
$\lambda(\vartheta-\epsilon)$ and therefore we can write an expression for 
$\varphi\vartheta$ as follows \cite{NHA}:
\begin{eqnarray}
\varphi\vartheta=\lambda\vartheta+k \lambda(\vartheta+\omega)+k^{\prime}
\lambda(\vartheta+\frac{\omega}{2}+\frac{\omega^{\prime}}{2})+k^{\prime\prime}
\lambda(\vartheta+\frac{3\omega}{2}+\frac{\omega^{\prime}}{2}) \nonumber \\
+\lambda(\vartheta+\epsilon_1)+\lambda(\vartheta-\epsilon_1)+
\lambda(\vartheta+\epsilon_2)+\lambda(\vartheta-\epsilon_2)+\cdots+
\lambda(\vartheta+\epsilon_n)+\lambda(\vartheta-\epsilon_n) \nonumber \\
\end{eqnarray}
where $k,k^{\prime},k^{\prime\prime}$ are equal to 0 or 1.

\subsubsection{Rational solution for $y=\psi(x)$, when $k=k^{\prime}=k^{\prime\prime}=0$}

In this particular case, we have
\begin{eqnarray}
\varphi\vartheta &=&\lambda\vartheta+\lambda(\vartheta+\epsilon_1)+\lambda(\vartheta-\epsilon_1)+
\lambda(\vartheta+\epsilon_2)+\lambda(\vartheta-\epsilon_2) \nonumber \\
 &+&\cdots+\lambda(\vartheta+\epsilon_n)+\lambda(\vartheta-\epsilon_n) 
\end{eqnarray}
or written in terms of $x$ 
\begin{equation}
\varphi\vartheta=x+2 x \sum_i \frac{\Delta \epsilon_i}{1-e^2c^2 \lambda^2 
\epsilon_i x^2} 
\end{equation}
Thus, the prime condition that $y$ is rational in $x$ is fulfilled. 
Let $\delta,\delta^{\prime},\eta,\eta^{\prime}$ be the values of $\vartheta$ 
that correspond to $y=1/c_1,-1/c_1,1/e_1,-1/e_1$. Thus, for $y=1/c_1$,
$1-c_1 y=0=\frac{g^{\prime}-c_1 f^{\prime}+(g-c_1f) \varphi \delta}
{g^{\prime}+g\varphi\delta}$ or 
\begin{equation}
1-c_1y=\frac{g^{\prime}-c_1 f^{\prime}}{\Sigma}\left[1-\frac{\varphi\vartheta}{\varphi\delta}\right]
\end{equation}
where $\Sigma:=g^{\prime}+g\varphi\vartheta$.
Similarly, we get \cite{NHA}
\begin{eqnarray}
1+c_1y&=&\frac{g^{\prime}+c_1 f^{\prime}}{\Sigma}\left[1-\frac{\varphi\vartheta}{\varphi\delta^{\prime}}\right] \nonumber \\
1-e_1y&=&\frac{g^{\prime}-e_1f^{\prime}}{\Sigma}\left[1-\frac{\varphi\vartheta}{\varphi\eta}\right] \nonumber \\
1+e_1y&=&\frac{g^{\prime}+e_1f^{\prime}}{\Sigma}\left[1-\frac{\varphi\vartheta}{\varphi\eta^{\prime}}\right] \nonumber \\
\end{eqnarray}
From the expression of $\vartheta$, one gets 
\begin{equation}
1-\frac{\varphi\vartheta}{\varphi\delta}=
\frac{1+A_1 x+A_2 x^2+\cdots A_{2n+1} x^{2n+1}}
{(1-e^2 c^2 \lambda^2 \epsilon_1 x^2)
 (1-e^2 c^2 \lambda^2 \epsilon_2 x^2) \cdots
 (1-e^2 c^2 \lambda^2 \epsilon_n x^2)}
\label{NielsN}
\end{equation}
For $\vartheta=\delta$ the right hand side of (\ref{NielsN}) must vanish  
and similarly for $\delta\pm \epsilon_1,\cdots \delta\pm \epsilon _n$ with 
arbitrary $\delta$. Thus we have that 
\begin{eqnarray}
1+A_1 x+\cdots +A_{2n+1} x^{2n+1} &=&\left(1-\frac{x}{\lambda \delta}\right)
\left(1-\frac{x}{\lambda(\delta+\epsilon_1)}\right)
\left(1-\frac{x}{\lambda(\delta-\epsilon_1)}\right) \nonumber \\
 & \times & \cdots  
\left(1-\frac{x}{\lambda(\delta+\epsilon_n)}\right)
\left(1-\frac{x}{\lambda(\delta-\epsilon_n)} \right) \nonumber \\
\end{eqnarray} 
Eq.(\ref{TRANSF}) can be written as
\begin{equation}
\sqrt{(1-c_1^2 y^2)(1-e_1^2 y^2)}=\frac{1}{C}\frac{dy}{dx}\sqrt{
(1-c^2 x^2)(1-e^2 x^2)}
\end{equation}
and we can see that the four functions $1\pm c_1 y,1\pm e_1y$ must vanish 
when $x=\pm \frac{1}{c},\pm \frac{1}{e}$ that is when $\vartheta=\pm 
\frac{\omega}{2},\pm \frac{\omega^{\prime}}{2}$.
Thus for $\delta=\frac{\omega}{2},\delta^{\prime}=-\frac{\omega}{2},
\eta=\frac{\omega^{\prime}}{2},\eta^{\prime}=-\frac{\omega^{\prime}}{2}
$ we get $g^{\prime}=c_1 f\varphi \left(\frac{\omega}{2}\right)=e_1 f
\varphi\left(\frac{\omega^{\prime}}{2}\right)$ and  \footnote{We have used the 
fact that Abel's function satisfies: $\varphi\left(\frac{-\omega}{2}\right)=
-\varphi\left(\frac{\omega}{2}\right),\varphi\left(\frac{-\omega^{\prime}}{2}\right)=
-\varphi\left(\frac{\omega^{\prime}}{2}\right)$.}
$f^{\prime}=\frac{g}{c_1}\varphi\left(\frac{\omega}{2}\right)=\frac{g}{e_1}
\varphi\left(\frac{\omega^{\prime}}{2}\right)$.
A solution of this system is 
\begin{equation}
g=f^{\prime}=0,\frac{f}{g^{\prime}}=\frac{1}{\kappa},c_1=\frac{\kappa}{
\varphi(\frac{\omega}{2})},e_1=\frac{\kappa}{\varphi(\frac{\omega^{\prime}}{2})}\end{equation}
and $\kappa$ is arbritrary.
Then 
\begin{equation}
y=\frac{1}{\kappa}\varphi \vartheta
\end{equation}
and \footnote{$\lambda\frac{\omega}{2}=1/c$.}
\begin{equation}
1-\frac{\varphi\vartheta}{\varphi(\frac{\omega}{2})}=\frac{1}{\iota}
(1-c x)\left\{1-\frac{x}{\lambda(\frac{\omega}{2}-\epsilon_1)}\right\}^2
\left\{1-\frac{x}{\lambda(\frac{\omega}{2}-\epsilon_2)}\right\}^2 \cdots
\left\{1-\frac{x}{\lambda(\frac{\omega}{2}-\epsilon_n)} \right\}^2
\end{equation}
where $\iota:=(1-e^2 c^2 \lambda^2 \epsilon_1 x^2)(1-e^2 c^2 
\lambda^2 \epsilon_2 x^2)\cdots (1-e^2 c^2 \lambda^2 \epsilon_n x^2)$.
Similar expressions are obtained for $1+\frac{\varphi\vartheta}{\varphi(\frac{\omega}{2})},1\pm \frac{\varphi\vartheta}{\varphi(\frac{\omega^{\prime}}{2})}$. Then 
$1-c_1^2 y^2 =(1-\frac{\varphi\vartheta}{\varphi\frac{\omega}{2}})
(1+\frac{\varphi\vartheta}{\varphi\frac{\omega}{2}})=(1-c^2 x^2)\frac{d^2}
{\iota^2},1-e_1^2y^2=(1-e^2 x^2)\frac{d^{\prime 2}}{\iota^2}$ or
\begin{equation}
\sqrt{(1-c_1^2 y^2)(1-e_1^2 y^2)}=\pm \frac{dd^{\prime}}{\iota^2}
\sqrt{(1-c^2 x^2)(1-e^2 x^2)}
\end{equation}
where $d:=\left(1-\frac{x^2}{\lambda^2 (\frac{\omega}{2}-
\epsilon_1)}\right)\cdots\left(1-\frac{x^2}{\lambda^2 (\frac{\omega}{2}-
\epsilon_n)}\right)$ and a similar expression for $d^{\prime}$ by 
substituting $\omega/2\rightarrow \omega^{\prime}/2$.
Then it is shown \cite{NHA} that 
$\frac{\iota^2 \frac{dy}{dx}}{dd^{\prime}}$ is 
a constant $C$ and Eq.(\ref{TRANSF}) is satisfied.

Two other expressions for $y$ are also provided \cite{NHA}
\begin{equation}
y=C \frac{x \left(1- \frac{x^2}{\lambda^2 \epsilon_1}\right)
\left(1- \frac{x^2}{\lambda^2 \epsilon_2}\right)\cdots
\left(1- \frac{x^2}{\lambda^2 \epsilon_n}\right)}{
(1-e^2c^2\lambda^2 \epsilon_1 x^2)
(1-e^2c^2\lambda^2 \epsilon_2 x^2)\cdots
(1-e^2c^2\lambda^2 \epsilon_n x^2)
}
\end{equation} 
and
\begin{equation}
y=\frac{1}{\kappa}\;(ec)^{2n}b\;\lambda\vartheta\;\lambda(\epsilon_1+
\vartheta)\lambda(\epsilon_1-\vartheta)\cdots 
\lambda(\epsilon_n+
\vartheta)\lambda(\epsilon_n-
\vartheta)
\end{equation}
with $b:=\lambda^2 \epsilon_1 \lambda^2 \epsilon_2\lambda^2 
\epsilon_3\cdots \lambda^2 \epsilon_n$.
For $\vartheta=\omega/2,\vartheta=\omega^{\prime}/2$, with corresponding 
values for $y=1/c_1$ and $1/e_1$ we obtain
\begin{eqnarray}
\frac{1}{c_1} &=& (-1)^n \frac{b}{\kappa}e^{2n}c^{2n-1}
\left[\lambda^(\frac{\omega}{2}-\epsilon_1)
\lambda^(\frac{\omega}{2}-\epsilon_2)\cdots \lambda(\frac{\omega}{2}-
\epsilon_n)\right]^2 \nonumber \\
\frac{1}{e_1} &=& (-1)^n 
\frac{b}{\kappa}e^{2n-1}c^{2n}\left[\lambda^(\frac{\omega^{\prime}}{2}-\epsilon_1)
\lambda^(\frac{\omega^{\prime}}{2}-\epsilon_2)\cdots \lambda(\frac{\omega^{\prime}}{2}-
\epsilon_n)\right]^2 \nonumber \\
\end{eqnarray}
As is discussed in \cite{NHA} (see also \cite{NHA1}) the transformation found by Jacobi in 
\cite{CGJAC},
corresponds to $\epsilon_1=\frac{2\omega}{2n+1},c=c_1=1$, and then 
$\epsilon_2=\frac{4\omega}{2n+1}=2\epsilon_1$,$\epsilon_n=n\epsilon_1$ 
and 
\begin{eqnarray}
y&=&\frac{\lambda \vartheta \lambda\left(\frac{2\omega}{2n+1}+\vartheta\right)
\lambda\left(\frac{2\omega}{2n+1}-\vartheta\right)\cdots
\lambda\left(\frac{2n\omega}{2n+1}+\vartheta\right)
\lambda\left(\frac{2n\omega}{2n+1}-\vartheta\right)}
{\left[\lambda\left(\frac{1}{2n+1}\frac{\omega}{2}\right)
\lambda\left(\frac{3}{2n+1}\frac{\omega}{2}\right)\cdots
\lambda\left(\frac{2n-1}{2n+1}\frac{\omega}{2}\right)
\right]^2} \nonumber \\
C&=&\left\{\frac{\lambda\left(\frac{\omega}{2n+1}\right)
\lambda \left(\frac{2\omega}{2n+1}\right)\cdots \lambda\left(\frac{n\omega}{2n+1}\right)}{\lambda\left(\frac{1}{2n+1}\frac{\omega}{2}\right)\lambda\left(
\frac{3}{2n+1}\frac{\omega}{2}\right)\cdots\lambda\left(\frac{2n-1}{2n+1}
\frac{\omega}{2}\right)}\right\} ^2 \nonumber \\
e_1&=&e^{2n+1}\left[\lambda\left(\frac{1}{2n+1}\frac{\omega}{2}\right)
\lambda\left(\frac{3}{2n+1}\frac{\omega}{2}\right)\cdots
\lambda\left(\frac{2n-1}{2n+1}\frac{\omega}{2}\right)
\right]^4 \nonumber \\
\frac{dy}{\sqrt{(1-y^2)(1-e_1^2 y^2)}}&=&\pm C \frac{dx}{\sqrt{(1-x^2)
(1-e^2 x^2)}}=\pm C d\vartheta \nonumber \\
\label{ODDN}
\end{eqnarray}

In addition in \cite{NHA},two more cases are discussed i) those  in which $k=0$ and $k^{\prime}$ or $k^{\prime\prime}$ is equal to 1 and ii) $k=1$. The general theorem for the first real transformation of arbitrary order $n$ is also 
stated \cite{NHA}.

In Jacobi's language the indefinite elliptic integral $x=\int \frac{dy}{
\sqrt{(1-y^2)(1-k^2 y^2)}}$ defines the elliptic 
function $y={\rm sn}(x,k)$.Thus, 
in Eq.(\ref{ODDN}), $x=\lambda \vartheta={\rm sn}(\vartheta,e)$ 
\footnote{
One can also switch to Legendre's notation by defining 
$y=\sin\psi$, $x=\sin \varphi^{\prime}$. Then the last line of 
(\ref{ODDN}) can be written as 
$\frac{d\psi}{\sqrt{1-e_1^2\sin^2 \psi}}=
C\frac{d \varphi^{\prime}}{\sqrt{1-e^2 \sin^2 \varphi^{\prime}}}$. For large $n$ the modulus $e_1$ becomes negligible 
and then $\psi=\sin^{-1}y=C\int \frac{d \varphi^{\prime}}{\sqrt{1-e^2\sin^2 
\varphi^{\prime}}}$}.

\subsection{Transforming the geodesic elliptic integrals into Abel's form}

The transformation
\begin{equation}
x\rightarrow e_3+\frac{(e_1-e_3)}{x^2}
\end{equation}
transforms the elliptic integral in Weierstra$\ss$ form into Abel's and Jacobi's form:
\begin{equation}
\int \frac{dx}{\sqrt{4(x-e_1)(x-e_2)(x-e_3)}}\rightarrow -\frac{1}{\sqrt{e_1-e_3}}\int \frac{dx}{\sqrt{(1-x^2)(1-k^2 x^2)}}
\end{equation}
with $k^2=\frac{e_2-e_3}{e_1-e_3}$.

Similarly, the quartic can be brought to Jacobi's form $y^2=(1-x^2)(1-k^2_1 x^2)$.



Indeed, as we saw in section 3 we have 
\begin{equation}
\frac{d\theta}{\sqrt{\Theta}}=-\frac{1}{2}\frac{dz}{\sqrt{z^3 \alpha-
z^2 (\alpha+\beta)+Q z}}=-\frac{1}{2}\frac{d\xi}{\sqrt{4 \xi^3-g_2 \xi-
g_3}}
\end{equation}
where 
\begin{eqnarray}
g_2 &=& \frac{1}{12}\left(a^2(1-E^2)+Q+L^2\right)^2-\frac{Q}{4}(a^2 (1-E^2)) \nonumber \\
g_3 &=& \frac{1}{216} \left(a^2(1-E^2)+Q+L^2\right)^3
-\frac{Q}{48} a^4(1-E^2)^2-
\frac{Q}{48} (a^2(1-E^2) (Q+L^2)) \nonumber \\
\end{eqnarray}
In this case the three roots of the cubic $4 \xi^3-g_2 \xi-g_3$ are given 
by the expressions
\begin{eqnarray}
e_1 &=&\frac{1}{24} \left(a^2(1-E^2)+L^2+Q+
3\sqrt{(a^2(-1+E^2)+Q)^2+2 a^2 L^2-2 a^2 E^2 L^2+L^4+2 L^2 Q}\right) \nonumber \\
e_2 &=& \frac{1}{24} \left(a^2(1-E^2)+L^2+Q-
3\sqrt{(a^2(-1+E^2)+Q)^2+2 a^2 L^2-2 a^2 E^2 L^2+L^4+2 L^2 Q}\right) \nonumber \\
e_3 &=& \frac{1}{12} \left(a^2 (E^2-1)-L^2-Q\right) \nonumber \\
\end{eqnarray}
Thus the Jacobi modulus $k^2=\frac{e_2-e_3}{e_1-e_3}$ is given by
\begin{equation}
k^2=\frac{-a^2 (-1+E^2)+L^2+Q-\sqrt{a^4 (-1+E^2)^2-2 a^2 (-1+E^2)(L^2-Q)+
(L^2+Q)^2}}{-a^2 (-1+E^2)+L^2+Q+\sqrt{a^4 (-1+E^2)^2-2 a^2 (-1+E^2)(L^2-Q)+
(L^2+Q)^2}}
\end{equation}
It has the correct limit for polar geodesics with $L=0$
\begin{equation}
k^2 (L=0)=\frac{a^2 (1-E^2)}{Q}
\end{equation}

Also 
\begin{equation}
\frac{1}{\sqrt{e_1-e_3}}=\frac{2 \sqrt{2}}{\sqrt{-a^2 (-1+E^2)+
L^2+Q+\sqrt{a^4(-1+E^2)^2-2 a^2(-1+E^2)(L^2-Q)+(L^2+Q)^2}}}
\end{equation}
and it also has the correct limit for polar geodesics 
\begin{equation}
\frac{1}{\sqrt{e_1-e_3}}=\frac{2}{\sqrt{Q}}
\end{equation}

Thus, we get 
\begin{equation}
\int \frac{d\theta}{\sqrt{\Theta}}=-\frac{1}{2}\frac{1}{\sqrt{e_1-e_3}}
\int \frac{d \xi^{\prime}}{\sqrt{(1-\xi^{\prime 2})(1-k^2 \xi^{\prime 2})}}
\end{equation}

Now we shall discuss the transformation of the right hand side of 
Eq.(\ref{getimelike}). 
Inspired by the work of Jacobi and Abel, Luchterhand obtained the following
transformation formula \cite{LUCH}:
\begin{equation}
\frac{\partial x}{M \sqrt{(1-x^2)(1-k_1^2x^2)}}=\frac{\partial y}{
\sqrt{(y-\alpha)(y-\beta)(y-\gamma)(y-\delta)}}
\end{equation}
where the Jacobi modulus $k_1$ and the coefficient $M$, are given in terms of 
the roots $\alpha,\beta,\gamma,\delta$ of the quartic, by the following 
expressions
\begin{equation}
k_1=\frac{\sqrt{(\alpha-\delta)(\beta-\gamma)}}{\sqrt{(\alpha-\gamma)(
\beta-\delta)}},\;\;\;M=\sqrt{(\alpha-\gamma)(\beta-\delta)}
\end{equation}
and the integration variables are related by
\begin{equation}
\frac{1-x}{1+x}=\frac{(\gamma-\delta)(y-\alpha)(y-\beta)}{(\alpha-\beta)
(y-\gamma)(y-\delta)}
\end{equation}
Also we assume that the roots $\alpha,\beta,\gamma,\delta$ 
of the quartic are real and are organized in
the following ascending order of magnitude: $\alpha>\beta>\gamma>\delta$.

At this stage, we have succeeded in transforming both sides of Eq.( \ref{getimelike}), into Abel's form. 
We can also provide a nice formula for $r$ in terms of the 
Jacobi's sinus amplitudinus function \footnote{We wright
$\int \frac{dr}{\sqrt{R}}=\int \frac{dr}{\sqrt{(r-\alpha)(r-\beta)(r-\gamma)(r-\delta)}}=\int \frac{\partial x}{M \sqrt{(1-x^2)(1-k_1^2 x^2)}}=
\int \frac{d\theta}{\sqrt{\Theta}}$.}
\begin{equation}
\frac{(\gamma-\delta)(r-\alpha)(r-\beta)}{(\alpha-\beta)(r-\gamma)
(r-\delta)}=\frac{1-{\rm sn}\left(M \int \frac{d\theta}{\sqrt{\Theta}},k_1\right)}{1+{\rm sn}\left(M\int \frac{d\theta}{\sqrt{\Theta}},k_1\right)}
\end{equation}

A systematic investigation of the phenomenological aspects of the above 
exact solution  like the periapsis precession will be reported elsewhere.

\section{Separability of Hamilton-Jacobi's differential equation in 
Kerr-(anti) de Sitter metric and derivation of geodesics.}

In the presence of the cosmological constant we proved (see Appendix A 
for details) the important 
result that the Hamilton-Jacobi differential equation 
can be solved by separation of variables. Thus in this case, the 
 characteristic function separates and takes the form
\begin{eqnarray}
  W&=&-Ect +L\phi+\int\frac{\sqrt{\left[Q+(L-a E)^2 \Xi^2-
\mu^2 a^2 \cos^2\theta\right]\Delta_{\theta}-\frac{\Xi^2(a E\sin^2\theta-L)^2}{\sin^2\theta}}}{\Delta_{\theta}}d\theta \nonumber \\
&+&\int\frac{\sqrt{\Xi^2\left[(r^2+a^2)E-aL\right]^2-\Delta_r(\mu^2r^2+Q+
\Xi^2(L-a E)^2)}}{\Delta_r}dr \nonumber 
\end{eqnarray}

By differentiating now with respect to constants of integration, $Q,L,E,\mu$, 
we obtain the following set of geodesic differential equations
\begin{eqnarray}
\int \frac{dr}{\sqrt{R^{\prime}}}=\int \frac{d\theta}{\sqrt{\Theta^{\prime}}}
\nonumber \\
\rho^2 \frac{d\phi}{d\lambda}=-\frac{\Xi^2}{\Delta_{\theta}\sin^2\theta}
\left(aE\sin^2\theta-L\right)+\frac{a\Xi^2}{\Delta_r}\left[(r^2+a^2)E-aL\right]
\nonumber \\
c\rho^2 \frac{dt}{d\lambda}=\frac{\Xi^2(r^2+a^2)\left[(r^2+a^2)E-aL\right]}{\Delta_r}-\frac{a\Xi^2 (aE \sin^2\theta-L)}{\Delta_{\theta}} \nonumber \\
\rho^2\frac{dr}{d\lambda}=\pm \sqrt{R^{\prime}} \nonumber \\
\rho^2\frac{d\theta}{d\lambda}=\pm \sqrt{\Theta^{\prime}} 
\label{LambdaGeo}
\end{eqnarray}
where 
\begin{eqnarray}
R^{\prime}&:=&\Xi^2 \left[(r^2+a^2)E-aL\right]^2-\Delta_r\left(\mu^2r^2+Q+
\Xi^2(L-aE)^2\right) \nonumber \\
\Theta^{\prime}&:=&\left[Q+(L-aE)^2 \Xi^2-\mu^2a^2 \cos^2\theta\right]\Delta_{\theta}-\Xi^2\frac{(aE \sin^2\theta-L)^2}{\sin^2\theta}
\label{LARTH}
\end{eqnarray}

The first line of Eq.(\ref{LambdaGeo}) is a differential equation that relates 
a {\em hyperelliptic} Abelian integral to an elliptic integral  which is the generalisation 
of the theory of transformation of elliptic functions discussed in section 
 5, in the case of non-zero cosmological constant. The mathematical treatment 
of such a relationship was first discussed by Abel in \cite{NHA2}.

\subsection{Exact solution of spherical polar orbits with a cosmological constant}

Using the second and the fifth line of Eq.(\ref{LambdaGeo}), for $L=0$ and 
assuming a constant value for $r$, 
we obtain 
\begin{eqnarray}
\frac{d\phi}{d\theta}&=&\frac{-\frac{\Xi^2 a E}{\Delta_{\theta}}+\frac{a \Xi^2 (
r^2+a^2)E}{\Delta_r}}{\sqrt{\Theta^{\prime}}} \nonumber \\
&=&\frac{-\Xi^2 E a+B \Delta_{\theta}}{\Delta_{\theta} \sqrt{\Theta^{\prime}}}
\end{eqnarray}
where $B:=\frac{a \Xi^2 (r^2+a^2)E}{\Delta_r}$. 

Similarly, using the third and fifth line we obtain
\begin{eqnarray}
\frac{cdt}{d\theta}&=&\frac{ \Xi^2 (r^2+a^2)^2 E}{\Delta_r\sqrt{\Theta^{\prime}}}-\frac{a^2 \Xi^2 E \sin^2\theta}{\Delta_{\theta} \sqrt{\Theta^{\prime}}} \nonumber \\
&=&\frac{\Gamma \Delta_{\theta}-a^2 \Xi^2 E \sin^2\theta}{\Delta_{\theta} \sqrt{\Theta^{\prime}}}
\end{eqnarray}
and $\Gamma:=\frac{\Xi^2 (r^2+a^2)^2 E}{\Delta_r}$.

Now using the variable $z=\cos^2 \theta$, we obtain the following system 
of integral equations:
\begin{eqnarray}
\phi&=&\int \frac{\Xi^2 a E/2}{\sqrt{f(z)}}dz+\int \frac{B(1+
\frac{a^2 \Lambda}{3}z)/(-2)}{\sqrt{f(z)}}dz \nonumber \\
ct&=&\int \frac{\frac{-\Gamma}{2} (1+\frac{a^2 \Lambda}{3}z)}
{\sqrt{f(z)}}dz-\int \frac{\frac{a^2}{-2} \Xi^2 E (1-z)dz}{\sqrt{f(z)}}
\label{ABTJAI}
\end{eqnarray}
or
\begin{eqnarray}
\phi&=&\int^z \frac{(\alpha_1+\beta_1z)dz}{\sqrt{f(z)}} \nonumber \\
c\;t&=&\int^z \frac{(\gamma_1+\delta_1 z)dz}{\sqrt{f(z)}} 
\end{eqnarray}
where $f(z)=z (1-z) (Q+z (Qa^2\frac{\Lambda}{3}+\Xi^3a^2 E^2-\mu^2a^2)+
z^2 (-\mu^2a^4\frac{\Lambda}{3}))(1+a^2 \frac{\Lambda}{3} z)^2$.
Also we have defined
\begin{eqnarray}
  \alpha_1&=&\Xi^2\frac{a E}{2}-\frac{1}{2}\frac{a\Xi^2(r^2+a^2)E}{\Delta_r}
\nonumber \\
\beta_1&=&-\frac{1}{2}\frac{a\Xi^2 (r^2+a^2)E}{\Delta_r}\frac{a^2\Lambda}{3}
\nonumber \\
\gamma_1&=&-\frac{1}{2}\frac{\Xi^2 (r^2+a^2)E}{\Delta_r}+\frac{a^2\Xi^2E}{2}
\nonumber \\
\delta_1&=&-\frac{a^2\Lambda}{6}\frac{\Xi^2(r^2+a^2)^2 E}{\Delta_r}-
\frac{a^2\Xi^2E}{2} 
\end{eqnarray}

Equation (\ref{ABTJAI}) is a system of equations of Abelian  
integrals, whose $inversion$ in principle, involves genus-2 
Abelian-Siegelsche modular 
functions.
Indeed, this system is a particular case of Jacobi's inversion problem of hyperelliptic 
Abelian integrals of genus 2 \cite{Abel}-\cite{BAKER} (see Appendix B 
for details). Then, one can express 
$z$ as a single valued 
genus two Abelian theta function with argumenents $t,\phi$. However, 
since the polynonial $f(z)$ of sixth degree posses a double root it may 
well be that the Abelian genus-2 theta function degenerates and the 
final result can be expressed in terms of genus-1 modular functions. 
This issue will be investigated in detail elsewhere.

\subsection{Equatorial geodesics including the contribution of the
cosmological constant}

The equatorial geodesics (i.e. $\theta=\pi/2,Q=0$), with a 
nonzero cosmological constant,  may be obtained by  Eq.(\ref{LambdaGeo}) for 
the particular values of $Q,\theta$.
The  characteristic function in this case, has the form 
\footnote{See appendix A for details.}

\begin{equation}
W=-Ect+\int\frac{\sqrt{R^{\prime}}}{\Delta_r}dr+L\phi
\end{equation}
and the geodesics are given by the expressions:

\begin{eqnarray}
\frac{dr}{\sqrt R^{\prime}}&=&\frac{d\lambda}{r^2}  \nonumber \\
r^2\frac{d\phi}{d\lambda} &=& \frac{a (1+\frac{1}{3}a^2\Lambda )^2(E(r^2
+a^2)-La) }{(1-\frac{\Lambda}{3}r^2)(r^2+a^2)-\frac{2 G M r}{c^2}}
+(L-a E)(1+\frac{1}{3}a^2 \Lambda )^2 \nonumber \\
c r^2\frac{dt}{d\lambda} &=&\frac{(1+\frac{1}{3}a^2\Lambda
)^2(r^2+a^2)\left[(r^2+a^2)E-a L\right]}{\Delta_r}
+(1+\frac{1}{3} a^2 \Lambda)^2 a (L-a E)  \nonumber \\
\label{equilambda}
\end{eqnarray}
where
\begin{equation}
R^{\prime}=(1+\frac{1}{3}a^2\Lambda )^2 \left[((r^2+a^2)E-a L)^2-
\Delta_r((L-a E)^2)\right]
\end{equation}
for null-geodesics and 
\begin{equation}
R^{\prime}=(1+\frac{1}{3}a^2\Lambda )^2 \left[((r^2+a^2)E-a L)^2-
\Delta_r((L-a E)^2)\right]-\Delta_r(\mu^2r^2)
\end{equation}
for time-like geodesics.

\subsection{Various limits of equatorial geodesic equations.}

For $\Lambda=0,a\not =0$ (Kerr limit) Eqs(\ref{equilambda}) reduces to
\begin{eqnarray}
\frac{dr}{\sqrt R}&=&\frac{d\lambda}{r^2}  \nonumber \\
r^2\frac{d\phi}{d\lambda} &=& \frac{a (E(r^2+a^2)-La) }{r^2+a^2-\frac{2 G M
r}{c^2}}+ (L-a E)
\nonumber \\
r^2 c\frac{dt}{d\lambda} &=&\frac{(r^2+a^2)\left[(r^2+a^2)E-a
L\right]}{\Delta}+a(L-aE)
\label{equi}
\end{eqnarray}
Similarly for $a=0,\Lambda\not= 0$ (Schwarzschild-de Sitter limit), Eqs(\ref{equilambda}) reduce to
\begin{eqnarray}
\frac{dr}{\sqrt {\left[r^4 E^2 -\left((1-\frac{\Lambda r^2 }{3})r^2-
\frac{2 G M r}{c^2}\right)[\mu^2
r^2+L^2]\right]}}&=&\frac{d\lambda}{r^2}\nonumber \\
r^2\frac{d\phi}{d\lambda} &=& L \nonumber \\
c dt &=&\frac{E}{1-\frac{\Lambda r^2}{3}-\frac{2 G M}{r c^2}}d\lambda
\nonumber \\
\label{SCDES}
\end{eqnarray}
The exact solution of (\ref{SCDES}) have been given in \cite{Mercury}.

\subsection{Circular geodesics with a cosmological constant}

In this section we shall study the circular equatorial geodesics and we shall prove 
the result:

\begin{eqnarray}
\frac{dt}{d\phi}&=&-
\frac{\Gamma_{03}^1 \pm \left [(\Gamma^1_{03})^2-\Gamma_{33}^1\Gamma_{00}^1\right]^{1/2}}{\Gamma^1_{00}} \nonumber \\ 
&=& \frac{ a}{c}\pm \frac{1}{c}\frac{\sqrt{ r}}{\sqrt{\left(\frac{ G M}{c^2 r^2}
-\frac{ \Lambda r}{3}\right)}} \nonumber \\
\label{circular}
\end{eqnarray}
where $\Gamma^i_{jk}$ denotes the Christoffel symbols.

The way to arrive to this result is to demand $\Gamma^1_{jk} u^j u^k=0$, 
together with the conditions:
\begin{eqnarray}
u^1=\frac{dr}{ds}=0  \nonumber \\
u^2=\frac{d\theta}{ds}=0 \nonumber \\
\end{eqnarray}

Alternatively since we have derived the general form for equatorial geodesics 
by requiring $u^1=0$ or equivalently that $R^{\prime}(r)=0$ (and its 
first derivative also vanishes) will lead us to a condition for eliminating the 
constants of integration and we should arrive in (\ref{circular}).

We also derive the following expression for $u^3=\frac{d\phi}{ds}$.
\begin{equation}
u^3=\frac{d\phi}{ds}=\left[g_{00}(\frac{dt}{d\phi})^2+2 g_{03}\frac{dt}{d\phi}+
g_{33}\right]^{-1/2}
\end{equation}

This is equivalent to the following equation
\begin{equation}
\frac{d\phi}{ds}=\frac{1}{\sqrt{\left [ \frac{c^2}{\Xi^2}(1-\frac{\Lambda}{3}(r^2+a^2)-
\frac{2 G M}{c^2 r})\left(\frac{dt}{d\phi}\right)^2+
2 \left\{\frac{ 2  G M a c}{c^2 r \Xi^2}+
\frac{c a (r^2+a^2)\frac{\Lambda}{3}}{(1+\frac{a^2 \Lambda}{3})^2}\right \}\frac{dt}{d\phi}-\left[\frac{(2 G M r)/c^2 a^2}{r^2 \Xi^2}+
\frac{(r^2+a^2) }{1+\frac{a^2 \Lambda}{3}}\right]\right]}}
\end{equation}

The above expression may be obtained from the metric
\begin{equation}
ds^2=g_{00}dt^2+g_{11}dr^2+g_{22}d\theta^2+g_{33}d\phi^2+2 g_{03}d\phi dt
\end{equation}

Some aspects of the equatorial circular geodesics have been investigated in 
\cite{KERNER}.

\section{Conclusions}

In this work, we have investigated the motion of a test particle in 
the gravitational field of Kerr space-time with and without the cosmological 
constant. The geodesic equations were derived by integrating the 
relativistic Hamilton-Jacobi partial differential equation by 
separation of variables.
As we saw in the main body of the paper, the structure of the exact 
solutions of the corresponding geodesic equations is quite rich and 
mathematically very interesting. The integration of the geodesic 
equations involves the inversion problem of Abel and Jacobi for 
hyperelliptic and elliptic Abelian integrals, as well as the theory 
of transformations of elliptic functions. 
As we saw, in the case of spherical polar orbits with a cosmological 
constant the solution involves directly the solution of Jacobi's inversion 
problem for a system of Abelian integrals.

Physically, the new types 
of relativistic motions lead to the interesting phenomenon of 
frame dragging of inertial frames. In this respect, we have applied 
the solution of spherical polar geodesics in Kerr space-time in 
two situations. First, by modelling the gravitational field of Earth 
by the Kerr geometry, we determined the longitudinal dragging of 
a satellite's polar spherical orbit around the Earth using as radius the 
semi-major axis of the polar orbit of GP-B mission launched in April 2004.
Secondly, by assuming the galactic centre is a rotating black hole, according 
to the interpretations of recent observations of the galactic 
centre \cite{GENZEL}, whose Kerr parameter is determined by experiment as 
for instance in Eq.(\ref{Galaxy}), we determined the longitudinal dragging of 
a star in spherical polar orbit around and close to the galactic centre. 
In principle, these predictions can be tested by observations.

A systematic investigation of the physical implications of the exact 
solutions of the geodesic equations for the various 
possible types of motion for the test particle, in the presence of the 
cosmological 
constant, will be reported elsewhere \cite{KRANIOTIS}.  
For instance, one can investigate the effect of the rotation 
of the Sun in the perihelion precession of Mercury and 
a comparison can be made with the cosmological constant contribution that 
has been calculated in \cite{Mercury}. 

It can also be applied in determining the  cosmological constant effect in rotating galaxies, and in particular in the velocity rotation curves. Assuming the galactic centre is a rotating black hole with the Kerr 
parameter measured as in eq.(\ref{Galaxy}), what is the effect of the cosmological constant in the motion of stars as a function of the distance from 
the galactic centre ?

Probing general relativity in the strong-field regime will certainly be 
one of the most exciting endeavours.

\section*{Acknowledgements} 
This work is supported by a Max Planck research fellowship at the Max 
Planck Institute for Physics in Munich.
The author also acknowledges the support from a research stipendium, 
during the early stages of this work, from the Max Planck Institute 
for Gravitational Physics (Albert Einstein Institute) at Golm.
We are gratefull to D. L$\rm{\ddot u}$st for useful discussions.

\appendix

\section{Separability of Hamilton-Jacobi equation in the presence of a cosmological constant}

In this appendix, we will prove that the Hamilton-Jacobi diffential equation 
preserves the important property of separability in the presence of the 
Cosmological constant, which has been used in deriving the geodesic equations
in section 6 from the characteristic function $W$.

The non-zero elements of the inverse metric are 
\begin{eqnarray}
g^{00}&=&\frac{\Xi^2\left[(r^2+a^2)^2\Xi-a^2\sin\theta^2[(r^2+a^2)\Xi-
\frac{2GMr}{c^2}]\right]}{c^2\Delta_r\Delta_{\theta}\rho^2} \nonumber \\
g^{03}&=&\frac{a\Xi^2\left(6 G Mr+c^2 \Lambda(a^2+r^2)\rho^2\right)}{
3c^2 \Delta_r\rho^2\Delta_{\theta}c} \nonumber \\
g^{11}&=&-\frac{\Delta_r}{\rho^2} \nonumber \\
g^{22}&=&-\frac{\Delta_{\theta}}{\rho^2} \nonumber \\
g^{33}&=&\Xi^2\left(\frac{a^2}{\Delta_r \rho^2}-\frac{1}{\rho^2\sin\theta^2
\Delta_{\theta}}\right) 
\label{invmetric}
\end{eqnarray}

The Hamilton-Jacobi differential equation takes the form:
\begin{eqnarray}
\Xi^2 &\times & \frac{ \left[(r^2+a^2)^2 \Xi-a^2 \sin^2\theta[(r^2+a^2)\Xi-\frac{2 G Mr}{c^2}]\right]E^2 \sin^2\theta}{\rho^2 \Delta_r\Delta_{\theta} \sin^2\theta}+
\frac{\Xi^2 (\sin^2\theta \Delta_{\theta}a^2-\Delta_r)L^2}{\rho^2\Delta_r\Delta_{\theta}\sin^2\theta} \nonumber \\
&+&\frac{2 a\Xi^2(\frac{2 G Mr}{c^2}+\frac{\Lambda}{3}\rho^2(a^2+r^2))(-EL)\sin^2\theta}{\rho^2 \Delta_r\Delta_{\theta}\sin^2\theta} \nonumber \\
&+&
\frac{-\Delta_r}{\rho^2}\left(\frac{\partial W_r}{\partial r}\right)^2+
\frac{-\Delta_{\theta}}{\rho^2}\left(\frac{\partial W_{\theta}}{\partial \theta}\right)^2+\mu^2=0
\end{eqnarray}

or
\begin{eqnarray}
\frac{\Xi^2}{\Delta_r}\left[(r^2+a^2)E-aL\right]^2-\Xi^2(L-aE)^2\frac{\Delta_r}{\Delta_r}+
\Xi^2 (L-aE)^2\frac{\Delta_{\theta}}{\Delta_{\theta}}-
\frac{\Xi^2}{\Delta_{\theta}}\frac{(aE\sin^2\theta-L)^2}{\sin^2\theta} +
\nonumber \\
+\mu^2 \rho^2+(-\Delta_r) \left(\frac{\partial W_r}{\partial r}\right)^2+
(-\Delta_{\theta})\left(\frac{\partial W_{\theta}}{\partial \theta}\right)^2=0
\nonumber \\
\end{eqnarray}

For equatorial geodesics the Hamilton-Jacobi differential equation takes 
the form
\begin{equation}
g^{00}(-E c)^2+g^{33}L^2+g^{11}\left(\frac{\partial W_r}{\partial r}\right)^2+
2g^{03}(-EcL)+\mu^2=0
\label{EquaHJ}
\end{equation}
Using eq.(\ref{invmetric}) with $\theta=\pi/2$ in (\ref{EquaHJ}) 
we obtain 
\begin{equation}
\frac{\partial W_r}{\partial r}=
\frac{\sqrt{\Xi^2\left[((r^2+a^2)E-aL)^2-\Delta_r(L-aE)^2\right]-\Delta_r\mu^2r^2}}{\Delta_r}
\end{equation}

\section{Definitions of genus-2 theta functions that solve Jacobi's inversion 
problem}

Riemann's theta function \cite{BerGeorgR} for genus $g$ is defined as follows:

\begin{equation}
\Theta(u):=\sum_{n_1,\cdots,n_g} e^{2\pi i u n+i \pi \Omega n^2}
\end{equation}
where $\Omega n^2:=\Omega_{11} n_1^2+\cdots 2 \Omega_{12}n_1 n_2 +\cdots$ 
 and $un:=u_1n_1+\cdots u_g n_g$. The symmetric $g\times g$ complex matrix 
$\Omega$ whose imaginary part is positive definite is a member of the 
set called Siegel upper-half-space denoted as ${\cal L}_{S_g}$. It is 
clearly the generalization of the ratio of half-periods $\tau$ in the genus $g=1$ case.    
For genus $g=2$ the Riemann theta function can be written in 
matrix form:

\begin{eqnarray}
\Theta(u,\Omega)&=&\sum_{{\bf{n}} \in Z^2} e^{\pi i {\bf ^{t} n}\Omega {\bf n}+
2\pi i {\bf ^{t} n} {\bf u}} \nonumber \\
&=& \sum_{n_1,n_2} e^{\pi i \left(\begin{array}{cc}n_1 & n_2\end{array}\right) \left(\begin{array}{cc}
\Omega_{11} & \Omega_{12} \\
\Omega_{12} &  \Omega_{22} \end{array}\right) \left(\begin{array}{c}
n_1 \\
n_2\end{array}\right) +2 \pi i \left(\begin{array}{cc} n_1 &n_2 \end{array}\right) \left(\begin{array}{c}
u_1 \\
u_2\end{array}\right)} \nonumber \\
\end{eqnarray}
Riemann's theta function with characteristics is defined by:
\begin{equation}
\Theta(u;q,q^{\prime}):=\sum_{n_1,\cdots,n_g}e^{2\pi i u(n+q^{\prime})+
i\pi \Omega (n+q^{\prime})^2+2 \pi i q(n+q^{\prime})}
\label{thechara}
\end{equation}
herein $q$ denotes the set of $g$ quantities $q_1,\cdots,q_g$ and $q^{\prime}$ 
denotes the set of $g$ quantities $q^{\prime}_{1},\cdots,q^{\prime}_g$.
Eq.(\ref{thechara}) can be rewritten in a suggestive matrix form:
\begin{equation}
\Theta \left [\begin{array}{c} q^{\prime} \\ q \end{array} \right](u,\Omega)=
\sum_{n \in Z^g} e^{\pi i ^{t} (n+q^{\prime}) \Omega (n+q^{\prime})+
2 \pi i ^{t} (n+q^{\prime})(u+q)}, \;\;\;\;q,q^{\prime} \in Q^g
\end{equation}

The theta functions whose quotients provide a solution to Abel-Jacobi's inversion 
problem are defined as follows \cite{BAKER}:
\begin{equation}
\theta(u;q,q^{\prime}):=\sum e^{au^2+2 hu(n+q^{\prime})+b(n+q^{\prime})^2+
2 i\pi q(n+q^{\prime})}
\end{equation}
where the summation extends to all positive and negative integer values of 
the $g$ integers $n_1,\cdots,n_g$, $a$ is any symmetrical matrix whatever 
of $g$ rows and columns, $h$ is any matrix whatever of $g$ rows and columns, 
in general not symmetrical, $b$ is any symmetrical matrix whatever of $g$ 
rows and columns, such that the real part of the quadratic form  $bm^2$ 
is necessarily negative for all real values of the quantities $m_1,\cdots,
m_g$, other than zero, and $q,q^{\prime}$ constitute the characteristics of 
the function. The matrix $b$ depends on $\frac{1}{2}g (g+1)$ independent 
constants; if we put $i\pi \Omega=b$ and denote the $g$-quantities 
$hu$ by $i\pi U$, we obtain the relation with Riemann's theta function:
\begin{equation}
\theta(u;q,q^{\prime})=e^{au^2}\Theta(U;q,q^{\prime})
\end{equation}

The dependence of genus-2 theta functions on two complex variables is
denoted by:  $\theta(u;q,q^{\prime})=\theta(u_1,u_2;q,q^{\prime})$,
the dependence on the Siegel moduli matrix $\Omega$ by:
$\theta(u_1,u_2,\Omega;q,q^{\prime})$.
To every half-period one can associate a set of characteristics.
For instance, the period $u^{a,a_1}=\frac{1}{2}\left(\begin{array}{cc}
1 & 0 \\
1 & 0 \end{array}\right)$ while $\theta(u)$ is a theta function of 
two variables with  zero
characteristics, i.e. $\theta(u)=\theta(u;0,0)=\theta \left
[\begin{array}{cc} 0 & 0 \\ 0 & 0 \end{array} \right](u,\Omega)$.
Also,
Weierstra$\ss$ had associated a symbol for each of the six odd theta
functions with characteristics 
and the ten even theta functions of genus two. For example,
$\theta(u)$ is associated with the Weierstra$\ss$ symbol 5 or
occasionaly the number appears as a subscript, i.e. $\theta(u)_5$.

Let the genus $g$  Riemann hyperelliptic surface be described by the 
equation:
\begin{equation}
y^2=4 (x-a_1)\cdots (x-a_g) (x-c) (x-c_1)\cdots (x-c_g)
\label{Riemann}
\end{equation}
For $g=2$ the above hyperelliptic Riemann algebraic equation reduces to:
\begin{equation}
y^2=4 (x-a_1) (x-a_2) (x-c) (x-c_1) (x-c_2)
\label{Riemann}
\end{equation}
where $a_1,a_2,c_,c_1,c_2$ denote the finite branch points of the surface.

The Jacobi's inversion problem involves  finding the  solutions,
for $x_i$ in terms of $u_i$,  
for the following system of equations of Abelian integrals \cite{BAKER}:
\begin{eqnarray}
u_1^{x_1,a_1} & +&\cdots + u_1^{x_g,a_g} \equiv  u_1 \nonumber \\
\vdots        &+ &\cdots +  \vdots \;\;\;\;\;\;\;\;\;\;\;\;\;  \vdots \nonumber \\
u_g^{x_1,a_1}  &+&\cdots + u_g^{x_g,a_g} \equiv u_g
\end{eqnarray}
where 
$u_1^{x,\mu}=\int_{\mu}^{x}\frac{dx}{y},
u_2^{x,\mu}=\int_{\mu}^{x}\frac{x dx}{y},
\cdots,u_g^{x,\mu}=\int_{\mu}^{x}\frac{x^{g-1} dx}{y}$.

For $g=2$ the above system of equations takes the form:
\begin{eqnarray}
\int_{a_1}^{x_1} \frac{dx}{y}+\int_{a_2}^{x_2} \frac{dx}{y}\equiv u_1 \nonumber \\
\int_{a_1}^{x_1} \frac{x\;dx}{y}+\int_{a_2}^{x_2} \frac{x\;dx}{y} \equiv u_2
\label{Umkehr}
\end{eqnarray}
where $u_1,u_2$ are arbitrary.
The solution  is given by the five equations \cite{BAKER} 
\begin{eqnarray}
\frac{ \theta^2(u|u^{b,a})}{
\theta^2(u)}&=&A(b-x_1)(b-x_2)\cdots(b-x_g) \nonumber \\
&=&A(b-x_1)(b-x_2) \nonumber \\
&=&\pm \frac{(b-x_1)(b-x_2)}{\sqrt{e^{\pi i P P^{\prime}}f^{\prime}(b)}};
\label{inveb}
\end{eqnarray}
where $f(x)=(x-a_1)(x-a_2)(x-c)(x-c_1)(x-c_2)$,
and $e^{\pi i P P^{\prime}}=\pm 1$ according as $u^{b,a}$ is an odd or even 
half-period. Also $b$ denotes a finite branch point and the branch
place $a$ being at infinity \cite{BAKER}.
The symbol $\theta(u|u^{b,a})$ denotes a genus 2 theta function with
characteristics: $\theta(u;q,q^{\prime})$ \cite{BAKER}, where $u,=(u_1,u_2)$, 
denotes two independent variables, see appendix A for further details.
From any $2$ of these equations, eq.(\ref{inveb}), the upper 
integration bounds $x_1,x_2$ 
of the system of differential equations eq.(\ref{Umkehr})
can be expressed as single valued 
functions of the arbitrary arguments $u_1,u_2$.
For instance,
\begin{equation}
x_1=a_1+\frac{1}{A_1 (x_2-a_1) } \frac{\theta^2(u|u^{a_1,a})}{\theta^2(u)}
\label{inve1}
\end{equation}
and 
\begin{eqnarray}
x_2&=&-\;\frac{\Bigl[(a_2-a_1)(a_2+a_1)+\frac{1}{A_1}\frac{\theta^2(u|u^{a_1,a})}{\theta^2(u)}-\frac{1}{A_2}\frac{\theta^2(u|u^{a_2,a})}{\theta^2(u)}\Bigr]}
{2 (a_1-a_2)} \nonumber \\
&\pm&\frac{\sqrt{\Bigl[(a_2-a_1)(a_2+a_1)+
\frac{1}{A_1}\frac{\theta^2(u|u^{a_1,a})}{\theta^2(u)}-\frac{1}{A_2}\frac{\theta^2(u|u^{a_2,a})}{\theta^2(u)}\Bigr]^2-
4(a_1-a_2)\eta}}{2(a_1-a_2)} \nonumber \\
\label{inve2}
\end{eqnarray}
where 
\begin{equation}
\eta:=a_2\;a_1(a_1-a_2)-\frac{a_2}{A_1}\frac{\theta^2(u|u^{a_1,a})}{\theta^2(u)}+\frac{a_1}{A_2}\frac{\theta^2(u|u^{a_2,a})}{\theta^2(u)}
\end{equation}
Also, $A_i=\pm \frac{1}{\sqrt{e^{\pi i P P^{\prime}}f^{\prime}(a_i)}}$.

The solution can be reexpressed in terms of generalized Weierstra$\ss$ functions:
\begin{equation}
x_k^{(1,2)}=\frac{\wp_{2,2}(u)\pm \sqrt{\wp^2_{2,2}(u)+4\wp_{2,1}(u)}}{2},
\;\;k=1,2
\end{equation}
where
\begin{equation}
\wp_{2,2}(u)=\frac{(a_1-a_2)(a_2+a_1)-\frac{1}{A_1}\frac{\theta^2(u|u^{a_1,a})}{\theta^2(u)}+\frac{1}{A_2}\frac{\theta^2(u|u^{a_2,a})}{\theta^2(u)}}{a_1-a_2}
\end{equation}
and 
\begin{equation}
\wp_{2,1}(u)=\frac{-a_1a_2(a_1-a_2)-\frac{a_1}{A_2}\frac{\theta^2(u|u^{a_2,a})}{\theta^2(u)}+\frac{a_2}{A_1}\frac{\theta^2(u|u^{a_1,a})}{\theta^2(u)}}{a_1-a_2}
\end{equation}
Thus, $x_1,x_2$, that solve Jacobi's inversion problem Eq.(\ref{Umkehr}), 
are solutions of a quadratic equation \cite{Jacobi,BAKER}
\begin{equation}
U x^2-U^{\prime} x+U^{\prime\prime}=0
\end{equation}
where $U,U^{\prime},U^{\prime\prime}$ are functions of $u_1,u_2$.
In the particular case that the coefficient of $x^5$ in the quintic 
polynomial is equal to 4, $U=1,U^{\prime}=\wp_{2,2}(u),U^{\prime\prime}=
\wp_{2,1}(u)$.

The matrix elements $h_{ij},\Omega_{ij}$ can be explicitly written in
terms of the half-periods $U_r^{x,a}$. For clarity,  $U_2^{e_4,e_3}=
\int_{e_3}^{e_4}xdx/y,U_1^{e_4,e_3}=\int_{e_3}^{e_4}dx/y$ etc. The roots have been arranged 
in ascending order of magnitude and are denoted by $e_{2g},e_{2g-1},\cdots,e_0,
g=2$, so that $e_{2i},e_{2i-1}$ are respectively, $c_{g-i+1}, 
a_{g-i+1}$ and $e_0$ is $c$.
For instance, the matrix element $h_{11}=\frac{U_2^{e_4,e_3}}{2(
U_1^{e_4,e_3}U_2^{e_2,e_1}-U_1^{e_2,e_1}U_2^{e_4,e_3})}\times \pi i$, while 
$\Omega_{11}=\frac{U_1^{e_1,e_0}U_2^{e_4,e_3}-U_2^{e_1,e_0}U_1^{e_4,e_3}}{
U_2^{e_2,e_1}U_1^{e_4,e_3}-U_1^{e_2,e_1}U_2^{e_4,e_3}}$.

Eq.(\ref{ABTJAI}) has the correct limit for vanishing cosmological constant.
Indeed, for $\Lambda=0$, $\beta_1=0$ and 
\begin{equation}
\phi=\int \frac{\alpha_1 dz}{\sqrt{z(1-z)(Q+z a^2(E^2-1))}}
\end{equation}
which can be transformed to Jacobi's form by the substitution $z=y^2$.
Then the solution is given by the Weierstra$\ss$ function of section 3.

\section{Transformation Theory of Elliptic functions and Modular equations}

One of the applications supplied by the transformation theory of Elliptic 
functions, 
which is of great importance in Number theory, are the modular equations 
described below \cite{NHA,NHA1,CGJAC}. 

For a rational solution of the differential equation
\begin{equation}
\frac{dy}{\sqrt{(1-y^2)(1-e_1^2y^2)}}=C\frac{dx}{\sqrt{(1-x^2)(1-e^2 x^2)}}
\label{TRANMOD}
\end{equation}
the necessary conditions among the periods 
\begin{eqnarray}
K(e_1)&=& a_0 C K(e)+a_1 C i K^{\prime}(e) \nonumber \\
iK(e_1)&=& b_0 C K(e)+ b_1 C i K^{\prime}(e) \nonumber \\
\end{eqnarray}
with the period ratios (moduli) of the associated modular theta functions 
being given by
\begin{equation}
\tau=\frac{b_0+b_1 \tau^{\prime}}{a_0+a_1\tau^{\prime}}
\end{equation}
are also sufficient,
when 
\begin{equation}
a_0 b_1-a_1 b_0=n
\label{modcorre}
\end{equation}
is a positive integer number.
The integer $n$ is called the degree of 
transformation.

Equation (\ref{modcorre}) for $a_0,b_1,a_1,b_0\in Z$ when 
viewed as the determinant of a matrix $\in GL(2,Z)$, sometimes is called a {\em modular correspondence of
level n}. 

It  can be shown that the {\em inequivalent reduced forms of modular 
correspondences} $$\left( \begin{array}{cc}
a_0 & a_1 \\ 
b_0 & b_1%
\end{array}%
\right),$$ are of the form 
  $$\left( \begin{array}{cc}
q & 0 \\ 
16\xi & q^{\prime}%
\end{array}%
\right)$$
where $q$ a positive part of $n$ represents, $q^{\prime}:=\frac{n}{q}$, 
and $0 \leq \xi \leq q^{\prime}-1$. For instance for $n=p$ a prime number, 
there are $p+1$ inequivalent reduced forms of the form \footnote{In 
a more familiar notation these classes of inequivalent reduced forms 
are $$\alpha_i=\left( \begin{array}{cc}
a & b \\ 
0 & d%
\end{array}%
\right), ad=n,\; (a,b,d)=1,\;0\leq b <d\; {\rm and}\; a,b,d\in Z$$.}
$$\left( \begin{array}{cc}
1 & 0 \\ 
0 & p%
\end{array}%
\right),\left( \begin{array}{cc}
1 & 0 \\ 
16 & p%
\end{array}%
\right), 
\left( \begin{array}{cc}
1 & 0 \\ 
16.2 & p%
\end{array}%
\right),
\cdots \left( \begin{array}{cc}
1 & 0 \\ 
16(p-1) & p%
\end{array}%
\right),
\left( \begin{array}{cc}
p & 0 \\ 
0 & 1
\end{array}
\right)$$

Also the multiplication factor $C$ in Eq.(\ref{TRANMOD}) is given by
\begin{equation}
C=\frac{1}{q}\frac{K(e_1)}{K(e)}
\label{MULTI}
\end{equation}
which  for a degree of transformation that is a prime number ($n=p$) is equal 
to $\frac{K(e_1)}{K(e)}$ or $(1/n)\frac{K(e_1)}{K(e)}$.
Eq.(\ref{MULTI}) can be reexpressed in terms of Jacobi theta functions as 
follows
\begin{equation}
C=\frac{1}{q}\frac{\vartheta_3^2(0,\tau)}{\vartheta_3^2(0,\frac
{q\tau-16\xi}{q^{\prime}})}
\end{equation}

The modular equations are equations relating the Jacobi modulus $e(p\tau)$ to 
$e(\tau)$ which are of the form 
\begin{equation}
F_p\left[\left(\frac{2}{p}\right)\sqrt[4]{e(\tau)},\sqrt[4]{e\left(\frac{\tau-16\xi}{p}\right)}\right]=0
\end{equation}
where $\left(\frac{2}{p}\right)$ denotes the Legendre symbol \footnote{
$\left(\frac{2}{a_0}\right)=e^{\frac{a_0^2-1}{8}i\pi}$.} .
Equivalently modular equations can be written in terms of 
the absolute modular invariant function $j(\tau)$, and relate the reduced 
absolute 
modular invariant $j^{*}$ to $j$
by polynomial equations of the form
\begin{equation}
\Phi_p(j^{*},j)=0
\end{equation}
where $j^{*}:=j. \alpha_i=j\left(\frac{a\tau+b}{d}\right)$
The explicit form of $\Phi_2(j^{*},j)=0$, has been given in \cite{Yui}.

An additional example considered in \cite{NHA1} is that of the differential 
equation 
\begin{equation}
\frac{dy}{\sqrt{(1-y^2)(1+e^2y^2)}}=C\sqrt{-1}\frac{dx}{\sqrt{(1-x^2)(1+e^2x^2)}}
\end{equation}
where $y=\pm i e^n x\frac{\left(\varphi^2(\frac{2\omega}{2n+1})-x^2\right)\cdots
\left(\varphi^2(\frac{n\omega}{2n+1})-x^2\right)}{(1+e^2\varphi^2(\frac{\omega}{2n+1})x^2)\cdots(1+e^2\varphi^2(\frac{n\omega}{2n+1})x^2)}$, while the modulus 
$e$ is given by $1=e^{n+1}(\varphi(\frac{\omega}{2(2n+1)})\cdots\varphi(
\frac{(2n-1)\omega}{2 (2n+1)}))^2$ and 
$C=\left(\frac{\varphi(\omega/(2n+1))\cdots\varphi(n\omega/(2n+1))}{\varphi(\frac{1}{2n+1}\frac{\omega}{2})\cdots
\varphi(\frac{2n-1}{2n+1}\frac{\omega}{2})}\right)^2\frac{1}{e}$. For $n=1$ it gives 
$e=\sqrt{3}+2,\varphi^2(\frac{\omega}{3})=2\sqrt{3}-3$. Also here 
$\varphi \vartheta=x$, where $\vartheta=\int\frac{dx}
{\sqrt{(1-x^2)(1+e^2x^2)}}$.

\section{Quartic and cubic for spherical orbits assuming $\Lambda=0$}

In the case of spherical orbits (polar with $L=0$, or non polar) and with 
the assumption of a vanishing cosmological constant, the test particle's 
radial coordinate will be stable at some value $r_f$ if $R(r)$ vanishes 
at $r=r_f$ and goes negative nearby. This will be the case if
\begin{eqnarray}
R(r_f)&=&0, \nonumber \\
\frac{dR}{dr}|r=r_f &=&0 \nonumber \\
\end{eqnarray}
and the second derivative of the polynomial $R(r)$ negative. 

These conditions result in the solution of a  quartic and a cubic algebraic 
equation. Let us describe first how we can solve the quartic equation:

For a quartic polynomial Ferrara (1545) solved the general equation of 
degree 4 :
\begin{equation}
P_4(x)=x^4-c_1 x^3+c_2 x^2-c_3 x+c_4=0
\end{equation}
as follows:
using the substitution $x \rightarrow x+ c_1/4 $, we can eliminate 
the cubic term  ($c_1=0$). 
The reduced equation is equivalent to 
\begin{equation}
(x^2+c_0)^2=y x^2+c_3 x +c_0^2-c_4
\end{equation}
where $c_0=\frac{1}{2}(c_2+y)$. The discriminant $\Delta$ of the right hand 
side (RHS) is:
\begin{equation}
\Delta=c_3^2-4 y (c_0^2-c_4)=c_3^2-4 y \left(\frac{1}{4}(c_2+y)^2-c_4\right)
\end{equation}
Its vanishing results in a cubic in $y$ to be solved.
Then the RHS has a double root $x_0=-\frac{c_3}{2 y}$, and 
$x^2+c_0=\pm \sqrt{y}(x-x_0)$ maybe solved by the extraction of one more 
square root.

As we described, the vanishing of the discriminant $\Delta=0$ results in 
a cubic equation:
\begin{equation}
-y^3-2 c_2 y^2+y (4 c_4-c_2^2)+c_3^2=0
\end{equation} 
or 
$y^3+2 c_2 y^2+y ( c_2^2-4 c_4)-c_3^2=0$. Under the substitution 
\begin{equation}
y=-4 \xi-\frac{2 c_2}{3}
\end{equation}
we get the cubic 
\begin{equation}
-64 \xi^3+\xi (\frac{4}{3} c_2^2+16 c_4)-\frac{2}{27}c_2^3-c_3^2+
\frac{8 c_2 c4}{3}
\end{equation}
or 
$\xi^3+\xi m-n=0$ with 
\begin{eqnarray}
m:=-\frac{m^{\prime}}{64},\;\; m^{\prime}=\frac{4}{3}c_2^2+16 c_4, \nonumber \\
n:=\frac{n^{\prime}}{64}\;\;,n^{\prime}=-\frac{2}{27}c_2^3-c_3^2+\frac{8 c_2 c_4}{3}
\end{eqnarray}
Then the roots are:

For polar orbits the coefficients $c_2,c_3,c_4$ in terms of the constants of 
integration $E,Q$, ($L=0$) and the Kerr parameter $a$ are given by
\begin{eqnarray}
c_3&=&\frac{-\frac{2 GM}{c^2}(Q+a^2 E^2)}{E^2-1}, \nonumber \\
c_2&=&a^2-\frac{Q}{E^2-1}, \nonumber \\
c_4&=&-\frac{Q a^2}{E^2-1}
\end{eqnarray}

\end{document}